\documentclass[aps,prc,preprint,superscriptaddress,showpacs]{revtex4}
\usepackage[dvips]{graphicx}
\usepackage{amsmath,amssymb}
\usepackage{ulem}

\newcommand{\beq}{\begin{equation}}
\newcommand{\eeq}{\end{equation}}
\newcommand{\bea}{\begin{eqnarray}}
\newcommand{\eea}{\end{eqnarray}}

\DeclareSymbolFont{boldletters}{OML}{cmm} {b}{it}
\DeclareSymbolFontAlphabet{\mathbit}{boldletters}
\DeclareMathSymbol{\alpha}{\mathalpha}{letters}{"0B}
\DeclareMathSymbol{\beta}{\mathalpha}{letters}{"0C}
\DeclareMathSymbol{\gamma}{\mathalpha}{letters}{"0D}
\DeclareMathSymbol{\delta}{\mathalpha}{letters}{"0E}
\DeclareMathSymbol{\epsilon}{\mathalpha}{letters}{"0F}
\DeclareMathSymbol{\zeta}{\mathalpha}{letters}{"10}
\DeclareMathSymbol{\eta}{\mathalpha}{letters}{"11}
\DeclareMathSymbol{\theta}{\mathalpha}{letters}{"12}
\DeclareMathSymbol{\iota}{\mathalpha}{letters}{"13}
\DeclareMathSymbol{\kappa}{\mathalpha}{letters}{"14}
\DeclareMathSymbol{\lambda}{\mathalpha}{letters}{"15}
\DeclareMathSymbol{\mu}{\mathalpha}{letters}{"16}
\DeclareMathSymbol{\nu}{\mathalpha}{letters}{"17}
\DeclareMathSymbol{\xi}{\mathalpha}{letters}{"18}
\DeclareMathSymbol{\pi}{\mathalpha}{letters}{"19}
\DeclareMathSymbol{\rho}{\mathalpha}{letters}{"1A}
\DeclareMathSymbol{\sigma}{\mathalpha}{letters}{"1B}
\DeclareMathSymbol{\tau}{\mathalpha}{letters}{"1C}
\DeclareMathSymbol{\upsilon}{\mathalpha}{letters}{"1D}
\DeclareMathSymbol{\phi}{\mathalpha}{letters}{"1E}
\DeclareMathSymbol{\chi}{\mathalpha}{letters}{"1F}
\DeclareMathSymbol{\psi}{\mathalpha}{letters}{"20}
\DeclareMathSymbol{\omega}{\mathalpha}{letters}{"21}
\DeclareMathSymbol{\varepsilon}{\mathalpha}{letters}{"22}
\DeclareMathSymbol{\vartheta}{\mathalpha}{letters}{"23}
\DeclareMathSymbol{\varpi}{\mathalpha}{letters}{"24}
\DeclareMathSymbol{\varrho}{\mathalpha}{letters}{"25}
\DeclareMathSymbol{\varsigma}{\mathalpha}{letters}{"26}
\DeclareMathSymbol{\varphi}{\mathalpha}{letters}{"27}
\DeclareMathSymbol{\Gamma}{\mathalpha}{letters}{"00}
\DeclareMathSymbol{\Delta}{\mathalpha}{letters}{"01}
\DeclareMathSymbol{\Theta}{\mathalpha}{letters}{"02}
\DeclareMathSymbol{\Lambda}{\mathalpha}{letters}{"03}
\DeclareMathSymbol{\Xi}{\mathalpha}{letters}{"04}
\DeclareMathSymbol{\Pi}{\mathalpha}{letters}{"05}
\DeclareMathSymbol{\Sigma}{\mathalpha}{letters}{"06}
\DeclareMathSymbol{\Upsilon}{\mathalpha}{letters}{"07}
\DeclareMathSymbol{\Phi}{\mathalpha}{letters}{"08}
\DeclareMathSymbol{\Psi}{\mathalpha}{letters}{"09}
\DeclareMathSymbol{\Omega}{\mathalpha}{letters}{"0A}

\begin{document}
\title{Roberge-Weiss phase transition and its endpoint}

\author{Hiroaki Kouno}
\email[]{kounoh@cc.saga-u.ac.jp}
\affiliation{Department of Physics, Saga University,
             Saga 840-8502, Japan}

\author{Yuji Sakai}
\email[]{sakai@phys.kyushu-u.ac.jp}
\affiliation{Department of Physics, Graduate School of Sciences, Kyushu University,
             Fukuoka 812-8581, Japan}

\author{Kouji Kashiwa}
\email[]{kashiwa@phys.kyushu-u.ac.jp}
\affiliation{Department of Physics, Graduate School of Sciences, Kyushu University,
             Fukuoka 812-8581, Japan}

\author{Masanobu Yahiro}
\email[]{yahiro@phys.kyushu-u.ac.jp}
\affiliation{Department of Physics, Graduate School of Sciences, Kyushu University,
             Fukuoka 812-8581, Japan}

\date{\today}

\begin{abstract}
The Roberge-Weiss (RW) phase transition in 
the imaginary chemical potential region is analyzed by 
the Polyakov-loop extended Nambu--Jona-Lasinio (PNJL) model. 
In the RW phase transition, the charge-conjugation symmetry 
is spontaneously broken, while the extended ${\mathbb Z}_{3}$ symmetry 
(the RW periodicity) is preserved. 
The RW transition is of second order at the endpoint. 
At the zero chemical potential,  
a crossover deconfinement transition appears 
as a remnant of the second-order RW phase transition at the endpoint, 
while the charge-conjugation symmetry is always preserved. 

\end{abstract}

\maketitle


\section{Introduction}
\label{sec:Introduction}

One of the most fascinating and essential subjects in hadron physics is 
to explore the phase diagram of quantum chromodynamics (QCD). 
QCD is a remarkable  theory in the sense that 
it is renormalizable and parameter free. 
The thermodynamics of QCD is well defined, nevertheless 
not clearly understood because of the nonperturbative nature. 
A powerful method of exploring the phase diagram is 
lattice QCD (LQCD) as the first-principle calculation, 
but it has the well known sign problem when the quark chemical potential 
($\mu$) is real; for example, see Ref.~\cite{Kogut} and references therein. 
Although several approaches such as the reweighting method~\cite{Fodor}, 
the Taylor expansion method~\cite{Allton} and 
the analytic continuation to the real chemical potential ($\mu_\mathrm{R}$) 
from the imaginary chemical potential 
($\mu_\mathrm{I}$)~\cite{FP,FP3,Elia1,Elia2,Chen34,Lomb,Chen,ERL} 
have been proposed, these are still far from perfection.  

So far the phase diagram in the $\mu_\mathrm{R}$ region has been analyzed by 
effective models such as the  Nambu--Jona-Lasinio (NJL) 
model~\cite{NJ1,AY,BR,Sca,Fujii,Fujii2,KKKN,Kashiwa,Sakaetal} and 
the Polyakov-loop extended Nambu--Jona-Lasinio (PNJL) 
model~\cite{Meisinger,Dumitru,Dumitru2005,Fukushima,Ghos,Megias,Ratti1,Ratti,Ciminale,
Rossner,Hansen,Sasaki,Schaefer,Costa,Kashiwa1,Fu,Abuki,
Sakai1,Fukushima2,Kashiwa4,Sakai2}. 
The NJL model describes the chiral symmetry breaking, but not 
the confinement mechanism. 
The PNJL model is designed \cite{Fukushima} 
to make it possible to treat both the mechanisms.  
It is reported that 
the confinement mechanism shifts 
the critical endpoint of the chiral phase transition
toward larger $T$ and smaller 
$\mu_\mathrm{R}$~\cite{Rossner,Kashiwa1,Fukushima2}.

When the chemical potential is imaginary, 
that is $\mu=i\mu_\mathrm{I}=iT\theta$, 
LQCD has no sign problem, so that 
LQCD data are available there~\cite{FP,FP3,Elia1,Elia2,Chen34,Lomb,Chen,ERL}. 
The first essential work on the $\mu_\mathrm{I}$ region was made
by Roberge and Weiss (RW)~\cite{RW}. 
They found that the thermodynamic potential 
$\Omega_{\rm QCD}(\theta)$ has a periodicity, 
$\Omega_{\rm QCD}(\theta)=\Omega_{\rm QCD}(\theta+2\pi k/3)$, 
for any integer $k$.  The RW periodicity was proven by showing that 
$\Omega_{\rm QCD}(\theta+2\pi k/3)$ is reduced to $\Omega_{\rm QCD}(\theta)$ 
with the ${\mathbb Z}_{3}$ transformation, 
\bea
q \to U q, \quad 
A_{\nu} \to UA_{\nu}U^{-1} - i/g (\partial_{\nu}U)U^{-1} \;, 
\label{Z3-transformation}
\eea
where $q$ is the quark field, $A_\nu$ is the gauge field and 
$U(x,\tau)$ are elements of SU(3) with 
$
U(x,1/T)=\exp(-2i \pi k/3)U(x,0).  
$
This means that $\Omega_{\rm QCD}(\theta)$ is invariant under the 
extended ${\mathbb Z}_{3}$ transformation~\cite{Sakai1}, 
\bea
&&\theta \to \theta + 2 \pi k/3, \nonumber \\  
&&q \to U q, \quad 
A_{\nu} \to UA_{\nu}U^{-1} - i/g (\partial_{\nu}U)U^{-1} \;. 
\label{EQ3}
\eea
The thermodynamic potential  $\Omega_{\rm QCD}(\theta)$ is transformed 
into $\Omega_{\rm QCD}(\theta+2\pi k/3)$ by the first 
transformation of Eq.~\eqref{EQ3} and it is transformed back into 
$\Omega_{\rm QCD}(\theta)$ by the ${\mathbb Z}_{3}$ transformation, 
that is, by 
the second and third equations of Eq.~\eqref{EQ3}. 
All quantities invariant under the extended ${\mathbb Z}_3$ 
transformation, such as the thermodynamic potential and 
the chiral condensate, keep the RW periodicity. 
Meanwhile, the Polyakov loop $\Phi$ is transformed 
as $\Phi \to \Phi e^{-i{2\pi k/3}}$ under the transformation \eqref{EQ3} 
and then does not have the RW periodicity. However, 
this problem can be solved by 
introducing the modified Polyakov loop $\Psi=\Phi e^{i\theta}$~\cite{Sakai1} 
that is invariant under the extended ${\mathbb Z}_3$ 
transformation. 
As an essential property, thus, 
QCD has the extended ${\mathbb Z}_{3}$ symmetry and it is realized as the RW periodicity in the $\mu_\mathrm{I}$ region. 

\begin{figure}[htbp]
\begin{center}
 \includegraphics[width=0.30\textwidth,angle=-90]{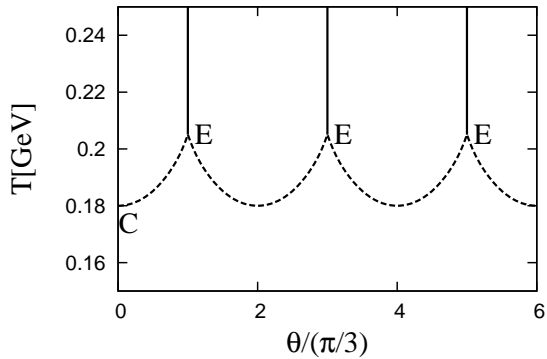}
 \end{center}
\caption{
Phase diagram on the $\theta$--$T$ plane predicted 
by the PNJL model. 
This diagram has a periodicity of $\theta=2\pi/3$.
The dashed curves represent crossover deconfinement phase transitions, 
while the solid lines represent the RW phase transition and its ${\mathbb Z}_{3}$ 
images. Points E are endpoints of the RW phase transitions. 
Point C is a pseudocritical transition point of the 
crossover deconfinement phase transition at $\theta=0$. 
}
\label{fig-PD-im}
\end{figure}

Among many effective models proposed so far 
the PNJL model is only a realistic effective model with 
both the extended ${\mathbb Z}_{3}$ symmetry 
and chiral symmetry~\cite{Sakai1,Kashiwa4,Sakai2}.  
As a result of this property, 
the PNJL model succeeds 
in reproducing the RW periodicity~\cite{Sakai1,Sakai2}. 
Figure~\ref{fig-PD-im} 
shows the two-flavor phase diagram in the 
$\theta$-$T$ plane predicted by the PNJL model; the details of 
the PNJL calculation will be described in Sec. \ref{sec:PNJL}. 
The dashed curves represent crossover deconfinement phase transitions, 
where the pseudocritical temperature at each $\theta$ is determined 
by the peak position of the Polyakov-loop susceptibility.

Roberge and Weiss also showed by using the perturbative and 
the strong coupling QCD 
that $\Omega_{\rm QCD}(\theta)$ is smooth 
at $\theta=\pi/3$ (mod $2\pi/3$) when $T$ is low, but not 
when $T$ is high~\cite{RW}. 
This indicates that there exists a phase transition above a 
temperature $T_{\rm E}$. 
This discontinuity of $d\Omega_{\rm QCD}(\theta)/d\theta$ 
is called the RW phase transition. 
The RW phase transition is known to be first order 
at $T>T_{\rm E}$, but the order at the endpoint $T=T_{\rm E}$ 
is not clarified yet.   
The presence of the first-order RW phase transition 
is confirmed by LQCD~\cite{FP,FP3,Elia1,Elia2,Chen34,Lomb,Chen}. 
In Fig.~\ref{fig-PD-im}, the solid line represents 
the RW phase transition predicted by the PNJL calculation. 
Point E is the endpoint of the RW phase transition. 
Obviously, this phase transition also has the RW periodicity. 
This success of the PNJL model suggests 
a possibility that the phase diagram 
in the $\mu_\mathrm{R}$ region is determined from 
the thermodynamics in the $\mu_\mathrm{I}$ region by 
using the PNJL model, the parameters of which are fitted to 
reproduce LQCD data in the $\mu_\mathrm{I}$ 
region. 
Actually an analysis along this line has been made very recently 
in Ref.~\cite{Sakai2}. 
Thus, deep understanding of the phase structure 
in the $\mu_\mathrm{I}$ region is important 
to determine the phase diagram in the $\mu_\mathrm{R}$ region.

The study of the $\mu_{\rm I}$ region has another important aspect. 
Roberge and Weiss found that at $T > T_{\rm E}$ 
three ${\mathbb Z}_{3}$ vacua come out alternatively 
as $\theta$ varies from 0 to $2\pi$ 
and showed that at $\theta=\pi/3$ (mod $2\pi/3$) a transition 
from one of ${\mathbb Z}_{3}$ vacua to another happens~\cite{RW}. 
This mechanism is an origin of the RW phase transition appearing at 
$\theta=\pi/3$ (mod $2\pi/3$). We call this mechanism the RW mechanism. 
At $\theta=0$, one of three vacua is selected. 
At $T<T_{\rm E}$, 
there is no ${\mathbb Z}_{3}$ vacua, and hence 
the RW mechanism does not take place.

In Fig.~\ref{fig-PD-im}, 
the dashed curve between points C and E represents a line of 
crossover deconfinement transition. It is natural to think 
that the behavior of the crossover deconfinement transition 
on the line and particularly at point C is 
influenced by 
the critical behavior of the RW phase transition at point E. 
The RW phase transition is studied in Ref.~\cite{ERL} 
by LQCD with four-flavor staggered fermions. 
The work suggests that 
the critical behavior of the RW phase transition at point E 
is essential to nonperturbative features of strongly coupled 
quark-gluon plasma (sQGP) appearing 
at $\mu =0$ and $T_{\rm C}< T < 3T_{\rm C}$, 
where $T_{\rm C}$ is a temperature of point C in Fig.~\ref{fig-PD-im}. 
Thus, the study on the RW phase transition is 
also important to understand properties of the crossover deconfinement 
transition and sQGP at $\mu =0$. 

In this paper, using the two-flavor PNJL model, 
we investigate properties 
of the RW mechanism 
and the RW phase transition as a consequence of the mechanism. 
Concretely, the following three points are argued. 
First, we show that in the RW phase transition the 
charge conjugation ($C$) symmetry is spontaneously broken, while 
the extended ${\mathbb Z}_{3}$ symmetry (the RW periodicity) is preserved. 
Second, we show that the RW phase transition  
is of second order at the endpoint $(\theta,T)=(\pi/3,T_{\rm E})$. 
In the $\mu_\mathrm{R}$ region, 
it is well known that 
the chiral phase transition is of second order 
at the critical endpoint~\cite{AY,Fujii,Fujii2}. 
Singular behaviors near the critical endpoints are 
different between the two second-order phase transitions. 
Third, we argue that the crossover deconfinement transition at $\mu=0$ is 
a remnant of the RW phase transition at the endpoint.

In Sec.~\ref{sec:PNJL}, 
the PNJL model and the extended ${\mathbb Z}_3$ symmetry 
are explained briefly. In Sec.~\ref{sec:results}, 
the RW phase transition and the RW mechanism 
are analyzed both analytically and 
numerically. 
Section~\ref{sec:summary} gives a summary.

\section{PNJL model}
\label{sec:PNJL}

\subsection{Model setting}

We consider the two-flavor PNJL Lagrangian with $\mu=i T \theta$,  
\begin{align}
 {\cal L}  =& {\bar q}(i \gamma_\nu D^\nu -m_0)q \notag\\
             &\hspace{3mm} + G_{\rm s}[({\bar q}q)^2 
                          +({\bar q}i\gamma_5 {\vec \tau}q)^2] 
              - {\cal U}(\Phi [A],{\Phi} [A]^*,T) ,
             \label{eq:E1}
\end{align}
where $m_0$ is the current quark mass, 
$D^\nu=\partial^\nu+iA^\nu-i\mu\delta^{\nu}_{0}$ and 
$A^\nu=\delta^{\nu}_{0}gA^0_a{\lambda^a\over{2}}$ 
with the gauge field $A^\nu_a$, 
the Gell-Mann matrix $\lambda_a$ and the gauge coupling $g$.
In the NJL sector, 
${\vec \tau}$ stands for the isospin matrix, and  
$G_{\rm s}$ denotes the coupling constant of the scalar-type 
four-quark interaction. 
The Polyakov potential ${\cal U}$, defined in Eq.~\eqref{eq:E13}, 
is a function of the Polyakov loop $\Phi$ and its Hermitian 
conjugate $\Phi^*$,
\begin{align}
\Phi      = {1\over{N_{\rm c}}}{\rm tr_{\rm c}} L,~~~~
\Phi^{*}  = {1\over{N_{\rm c}}} {\rm tr_{\rm c}}L^\dag ,
\end{align}
with
\begin{align}
L({\bf x}) = {\cal P} \exp\Bigl[
                {i\int^\beta_0 d \tau A_4({\bf x},\tau)}\Bigr],
\end{align}
where ${\cal P}$ is the path ordering, $A_4 = iA_0 $ and $N_{\rm c}=3$. 
In the PNJL model, $\Phi$ and $\Phi^{*}$ are treated as classical variables. 
We also denote $\Phi$ by $Re^{i\phi}$, where $R$ and $\phi$ are the absolute value  and the phase of $\Phi$, respectively. 
In the chiral limit ($m_0=0$), 
the Lagrangian density has the exact 
$SU(2)_{\rm L} \times SU(2)_{\rm R}
\times U(1)_{\rm v} \times SU(3)_{\rm c}$  symmetry.

Using the mean field approximation (MFA), 
one can obtain the thermodynamic potential per unit volume~\cite{Fukushima,Ratti1},
\begin{align}
\Omega =& -2 N_f \int \frac{d^3{\rm p}}{(2\pi)^3}
         \Bigl[ 3 E ({\rm p}) 
        + \frac{1}{\beta}
           \ln~(1 + F)
        \nonumber\\
        &+ \frac{1}{\beta} 
           \ln~(1 + F^{*})
	      \Bigl]
        +G_{\rm s}\sigma^2+{\cal U}  
\label{eq:E12} 
\end{align}
with 
\bea
F=3(\Phi+\Phi^{*} e^{-\beta E({\bf p})+i\theta }) 
           e^{-\beta E({\bf p})+i\theta}+ e^{-3\beta E({\bf p})+3i\theta} ,
\eea
where $N_f=2$, $\sigma =\langle \bar{q}q\rangle$ is the chiral condensate, 
$\beta=1/T$ and $E({\rm p})=\sqrt{{\bf p}^2+M^2}$ 
with the effective quark mass $M=m_0-2G_{\rm s}\sigma$. 
We use ${\cal U}$ of Refs.~\cite{Fukushima,Fukushima2} that has a 
strong coupling inspired form, 
\begin{align}
&{\cal U} = -bT \Bigl[54e^{-a/T}  \Phi {\Phi}^* 
+ \ln(1+G)\Bigr] , 
\label{eq:E13}
\end{align}
with 
\begin{align}
&G=- 6{\Phi\Phi^*}  + 4(\Phi^3+{\Phi^*}^3) - 3(\Phi\Phi^*)^2. \label{eq:G}
\end{align}
The constant parameter $a$ is taken to be 664MeV 
so as to reproduce the LQCD result on pure gauge that 
the first-order deconfinement phase transition takes place at 
$T=T_0=270$MeV~\cite{Fukushima,Fukushima2}.

The vacuum term (the first term of the right-hand side of Eq.~\eqref{eq:E12}) 
$\Omega^{\rm vac}$ diverges. 
It is then regularized 
by the three-dimensional momentum cutoff $\Lambda$: 
\begin{equation}
\int \frac{d^3{\bf p}}{(2 \pi)^3}\to 
{1\over{2\pi^2}} \int_0^\Lambda dp p^2. 
\label{eq:E15}
\end{equation}
Following Ref.~\cite{Fukushima}, we regularize the vacuum part, but not 
the thermal part. 
Even if the thermal part is regularized, it does not change the present 
result much unless $T$ is much larger than $T_{\rm E}$.

The parameter $b$ in ${\cal U}$ means a mixing strength 
between the chiral and deconfinement phase transitions, 
and is chosen to be 0.015$\Lambda^3$ 
to reproduce the two-flavor LQCD data in which 
a crossover deconfinement transition occurs around $T_{\rm C}\simeq 180$MeV; 
in ref.~\cite{Fukushima2}, $b$ is taken to be 0.03$\Lambda^3$ 
to reproduce $T_{\rm C}\simeq 200$MeV in the three-flavor case.

Hence, the present model has three parameters 
$m_0$, $\Lambda$, $G_{\rm s}$ in the NJL sector. 
Following Ref.~\cite{Kashiwa}, we use $m_0=5.5$~MeV, $\Lambda =0.6315$~GeV,  
and $G_{\rm s}=5.498$~GeV$^{-2}$ that reproduce 
the pion decay constant $f_{\pi}=93.3$MeV and the pion mass $M_{\pi}=138$MeV. 

In the $\mu_{\rm I}$ case, $\Omega$  and $\sigma$ are real, while 
$\Phi^*$ is the complex conjugate to $\Phi$~\cite{Sakai1}. 
Variables, $X=\Phi$, ${\Phi}^*$ and $\sigma$,  
satisfy the stationary conditions, 
\bea
\partial \Omega/\partial X=0.
\label{eq:SC}
\eea
The thermodynamic potential $\Omega(\theta)$ at each $\theta$ is then 
obtained by inserting the solutions $X(\theta)$ in Eq.~\eqref{eq:E12}. 
The thermodynamic potential $\Omega(\theta)$ thus obtained does not give 
the global minimum of $\Omega$ necessarily. 
Actually in the case of high $T$, 
there exist local minima (unstable solutions) in addition to the global minimum (the stable ground-state solution), as shown later with numerical calculations.

\subsection{Extended ${\mathbb Z}_3$ symmetry}


The thermodynamic potential $\Omega$ of Eq. (\ref{eq:E12}) is 
invariant under the extended ${\mathbb Z}_3$ transformation, 
\begin{align}
&e^{\pm i \theta} \to e^{\pm i \theta} e^{\pm i{2\pi k/3}},
\nonumber \\ 
&\Phi(\theta)  \to \Phi(\theta) e^{-i{2\pi k/3}}, 
\quad 
\Phi(\theta)^{*} \to \Phi(\theta)^{*} e^{i{2\pi k/3}} .
\label{eq:K2}
\end{align}
This is easily understood by introducing the modified Polyakov loop 
$\Psi \equiv e^{i\theta}\Phi$ and 
$\Psi^{*} \equiv e^{-i\theta}\Phi^{*}$ 
invariant under the transformation (\ref{eq:K2}). 
The extended ${\mathbb Z}_3$ transformation is then 
rewritten as
\begin{align}
e^{\pm i \theta} \to e^{\pm i \theta} e^{\pm i{2\pi k/3}}, \quad
\Psi(\theta) \to \Psi(\theta), \quad 
\Psi(\theta)^{*} \to \Psi(\theta)^{*} ,
\label{eq:K2'}
\end{align}
and $\Omega$ as 
\begin{align}
\Omega = \Omega^{\rm vac} + \Omega^{\rm ther} + {\cal U} ,
\label{eq:K3} 
\end{align}
with 
\begin{align}
\Omega^{\rm vac} &= -2 N_f \int \frac{d^3{\rm p}}{(2\pi)^3} 3 E ({\rm p}) 
          + G_{\rm s}\sigma^2 ,
\label{eq:omega-vac} \\
\Omega^{\rm ther} &= - \frac{2 N_f}{\beta} \int \frac{d^3{\rm p}}{(2\pi)^3}
         \Bigl[ \ln(1 + F) + \ln(1 + F^{*}) \Bigr] ,  
\label{eq:omega-log} \\
{\cal U} &= -bT \Bigl[54e^{-a/T} \Psi{\Psi}^*
            + \ln(1+G)\Bigr] ,
\label{eq:calU}
\end{align}
where 
\bea
F&=3\Psi e^{-\beta E({\bf p})}
         + 3\Psi^{*}e^{-2\beta E({\bf p})}e^{3i\theta}
          + e^{-3\beta E({\bf p})}e^{3i\theta},	~~~~\\
G&=-6{\Psi\Psi^*} 
            + 4(\Psi^3 e^{-3i\theta}+{\Psi^{*}}^3 e^{3i\theta})
            - 3(\Psi\Psi^*)^2  . ~~~~~~~~  
\eea
Obviously, $\Omega$ is extended ${\mathbb Z}_3$ invariant, since 
it is a function of only extended ${\mathbb Z}_3$ invariant 
quantities, $e^{3i\theta}$ and $X=\sigma$, $\Psi$ and $\Psi^{*}$. 
The explicit $\theta$ dependence appears only 
through the factor $e^{3i\theta}$ in Eq.~\eqref{eq:K3}. Hence, 
if the solutions $\{ X \}$ are uniquely given, 
they have ${X}={X}(e^{3i\theta})$. 
Inserting the solutions back into Eq.~\eqref{eq:K3}, one can see that 
$\Omega=\Omega(e^{3i\theta})$. 
Thus, $\Omega(\theta)$ and ${X}(\theta)$ have 
the RW periodicity, $\Omega(\theta)=\Omega(\theta+2\pi k/3)$ and 
${X}(\theta)={X}(\theta+2\pi k/3)$, 
because they are extended ${\mathbb Z}_3$ invariant. 
However, the situation is more complicated at high $T$, since 
three sets of solutions, $X_k$ ($k=0, \pm 1$), are given; this will be 
discussed in Sec.~\ref{sec:RW-mechanism}.

\section{Analytic and numerical results}
\label{sec:results}

\subsection{RW phase transition}
\label{sec:RW-phase-transition}

Under the charge conjugation ($C$), the Polyakov loop and the chemical 
potential $\mu$ are  
transformed as $\Phi \to \Phi^{*}$ and $\mu \to -\mu$, respectively; 
for example, see Refs. \cite{Dumitru2005, Fukushima2007}. 
This indicates that the modified Polyakov loop $\Psi=\Phi e^{i \theta}$ is 
also transformed as $\Psi \to \Psi^{*}$. 
As shown in Eqs.~\eqref{eq:omega-log} and \eqref{eq:calU}, 
$\Omega$ is invariant under the $C$ transformation. 
In other words, $\Omega$ is invariant under the transformation 
$\theta \to -\theta$, if $\Psi$ is replaced by $\Psi^{*}$. 
This means that the solutions $X(\theta)$ 
of the stationary conditions \eqref{eq:SC} satisfy 
\beq
\Psi(\theta)=\Psi^{*}(-\theta), \quad \sigma(\theta)=\sigma(-\theta). 
\label{Psi}
\eeq
Equation \eqref{Psi} indicates that the chiral condensate $\sigma$, 
the absolute value $|\Psi|$ and the real part Re$[\Psi]$ are $\theta$-even, 
while the phase $\psi={\rm arg}(\Psi)=\phi +\theta$ and the imaginary part Im$[\Psi]$ are 
$\theta$-odd. Inserting the solutions $X(\theta)$ into Eq.~\eqref{eq:K3}, 
one can see that 
$\Omega(\theta)$ is $\theta$-even. Hence, the derivative 
$d\Omega(\theta)/d\theta$ and the quark number density 
$n=-d\Omega/d\mu=-d\Omega/d(iT\theta)$ are 
$\theta$-odd quantities with the RW periodicity. 
Such $\theta$-odd quantities $O(\theta)$ with the RW periodicity 
satisfy 
\beq 
\lim_{\epsilon \to +0} O(\theta - \epsilon)
=- \lim_{\epsilon \to +0}O(\theta +\epsilon)
\label{odd}
\eeq 
at $\theta=\pi/3$ (mod $2\pi/3$), 
because 
$
O(\pi/3 - \epsilon)=-O(-\pi/3 + \epsilon)=-O(\pi/3 + \epsilon) .
$
Thus, $O(\theta)$ is discontinuous when 
$\lim_{\epsilon \to +0}O(\theta + \epsilon)$ is finite. 
This discontinuity is called the 
RW phase transition, and realized in the high $T$ region, 
$T>T_{\rm E}=205$~MeV, as shown below.

\begin{figure}[htbp]
\begin{center}
 \includegraphics[width=0.3\textwidth,angle=-90]{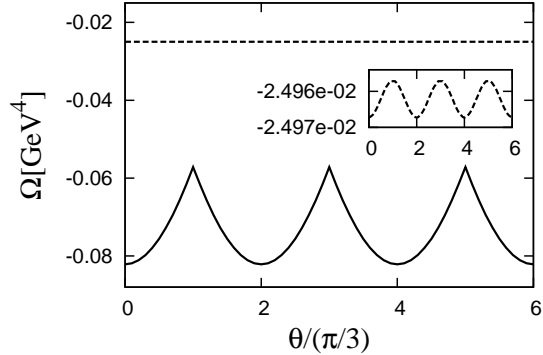} 
\end{center}
\caption{ 
Thermodynamic potential $\Omega$ as a function of $\theta$. 
The solid curve represents a result of the case of $T=400$~MeV, and 
the dashed one corresponds to that of $T=150$~MeV. 
The inset represents the $\theta$-dependence of $\Omega$ 
at $T=150$~MeV in a small scale. }
\label{theta_Omega_stable}
\end{figure}

Figure~\ref{theta_Omega_stable} shows $\theta$-dependence 
of $\Omega(\theta)$ calculated with the PNJL model. 
At $T=400$~MeV belonging to the the high-$T$ region $T>T_{\rm E}$, 
the solid curve has a cusp at $\theta=\pi/3$ (mod $2\pi/3$). 
Thus, $d\Omega(\theta)/d\theta$ is discontinuous there. 
At $T=150$~MeV belonging to the low-$T$ region $T<T_{\rm E}$, 
meanwhile, the dashed curve is smooth everywhere and 
has the RW periodicity, as shown by the inset.  
In this low-$T$ case, $\theta$-dependence of $\Omega$ is very weak. 
In the zero-$T$ limit, furthermore, $\Omega$ has no $\theta$-dependence, 
since 
\begin{align}
\Omega|_{T=0} \equiv \lim_{T \to 0} \Omega
 = -6N_f \int \frac{d^3{\rm p}}{(2\pi)^3}
           E ({\bf p}) + G_{\rm s}\sigma^2, 
\label{eq:E12-2}
\end{align}
where $\sigma$ is obtained by the stationary condition 
$\partial (\Omega|_{T=0})/\partial \sigma=0$. 

The solution $\Omega(\theta)$ is transformed by the charge conjugation $C$ as 
$\Omega(\theta) \to \Omega(-\theta)$. 
When the solution $\Omega(\theta)$ with $\theta$ fixed is considered, 
$C$ is a symmetry of it 
only at $\theta=0$ and $\pi$; 
note that $\theta=\pi$ is identical with $\theta=-\pi$. 
The $\theta$-odd quantity $O(\theta)$ such as 
$\psi$ and $n$ is transformed by $C$ as $O(\pi) \to -O(-\pi)=-O(\pi)$ and
hence not $C$-invariant at $\theta=\pi$. 
When $T<T_{\rm E}$, nevertheless, $O(\theta)$ is 
a smooth function of $\theta$, so that it is zero at $\theta=\pi$ 
because of Eq. \eqref{odd}. Thus, we can regard $O(\theta)$ 
as an order parameter of the $C$ symmetry and then use $\psi$ for this purpose. 

\begin{figure}[htbp]
\begin{center}
 \includegraphics[width=0.28\textwidth,angle=-90]{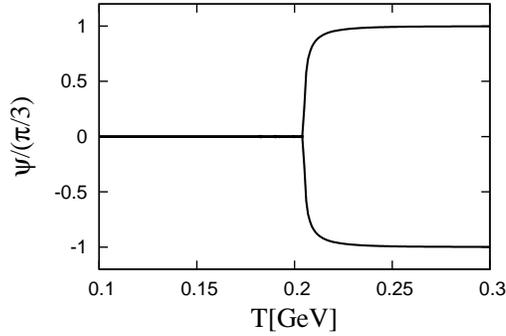}
\end{center}
\caption{Phase $\psi$ of the modified Polyakov loop as a function of $T$ 
in the case of $\theta =\pi$. 
}
\label{T-phi}
\end{figure}

Figure~\ref{T-phi} shows the $T$ dependence of $\psi$ at $\theta=\pi$. 
$\psi$ is zero at $T\leq T_{\rm E}=205$~MeV, while it is finite 
at $T>T_{\rm E}$. 
Thus, the $C$ symmetry is spontaneously broken 
above $T_{\rm E}$, although it is preserved below $T_{\rm E}$. 
Same phase transition takes place at $\theta=\pi/3$ and 
$5\pi/3$ as a consequence of the RW periodicity. 
At $T=T_{\rm E}$, $d\psi/dT$ is discontinuous, implying 
that the RW phase transition is of second order there. 
This result is consistent with 
the speculation of de Forcrand and Philipsen~\cite{FP} based on 
LQCD. 
Further discussion on this point will be made in subsection~\ref{sec:order}.

\begin{figure}[htbp]
\begin{center}
 \includegraphics[width=0.28\textwidth,angle=-90]{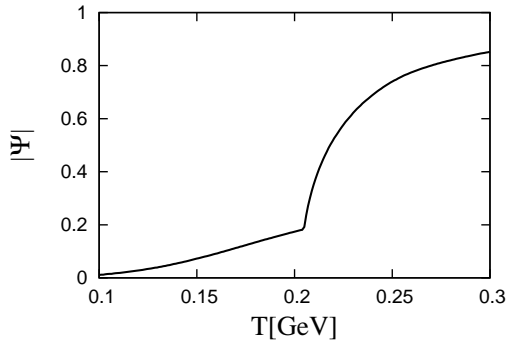}
\end{center}
\caption{
The absolute value of the modified Polyakov loop 
as a function of $T$ in the case of $\theta =\pi$. 
}
\label{FPsi}
\end{figure}

Figure~\ref{FPsi} shows $T$-dependence of 
the modulus $|\Psi|$ at $\theta=\pi$. 
The derivative $d|\Psi|/dT$ is also discontinuous at $T=T_{\rm E}$. 
This also implies that the RW phase transition is of second order 
at the endpoint $(\theta,T)=(\pi/3~{\rm mod}~2\pi/3,T_{\rm E})$, 
although $|\Psi|$ is not zero at $T< T_{\rm E}$ and then 
not an exact order parameter of this phase transition.

\begin{figure}[htbp]
\begin{center}
 \includegraphics[width=0.35\textwidth,angle=-90]{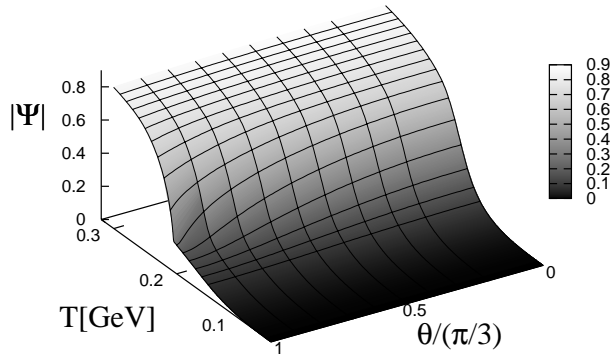}
\end{center}
\caption{
The absolute value of the modified Polyakov loop 
as a function of $T$ and $\theta$. }
\label{Psi_all}
\end{figure}

Figure~\ref{Psi_all} presents the modulus $|\Psi|$ as a function of $T$ and 
$\theta$. 
The second-order phase transition appearing at $\theta =\pi /3$ 
becomes crossover as $\theta$ decreases from $\pi /3$ to 0. 
Thus, the crossover deconfinement transition at $\theta=0$, shown by 
a rapid change of $|\Psi |$ with an increase of $T$, is a remnant of 
the second-order RW phase transition at $\theta =\pi/3$.

The PNJL analyses mentioned above are summarized as the phase diagram 
in Fig.~\ref{fig-PD-im}. 
The temperature of point C (the pseudocritical temperature of the 
deconfinement phase transition at $\theta=0$) is $T_{\rm C}=179$~MeV, and 
temperature of point E (the endpoint temperature 
of the RW phase transition) is $T_{\rm E}=1.15T_{\rm C}=205$~MeV. 
Thus, $T_{\rm E}$ is higher than $T_{\rm C}$. 

\subsection{RW mechanism}
\label{sec:RW-mechanism}

\begin{figure}[htbp]
\begin{center}
 \includegraphics[width=0.4\textwidth]{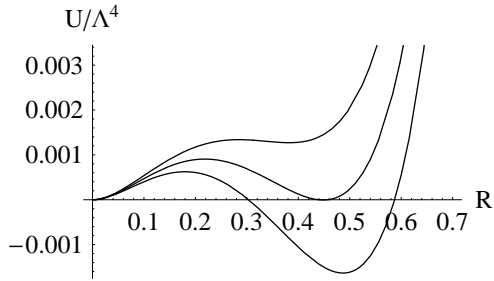}
\end{center}
\caption{
$R$ dependence of ${\cal U}/\Lambda^4$ at $\phi=0$ mod $2\pi/3$. 
Three solid curves correspond to the cases of $T/T_0=0.99, 1, 1.01$, 
respectively, 
from the top to the bottom. 
}
\label{pure-gauge-U}
\end{figure}

We start with the SU(3) pure gauge system. For $\Phi=R e^{i\phi}$, 
the Polyakov Potential ${\cal U}$ is obtained as 
\begin{align}
{\cal U}=&-bT
\Bigl[
54e^{-a/T}R^2
\notag\\
&+\log{(1-6R^2+8R^3\cos{3\phi}-3R^4)}
\Bigr] . 
\end{align}
The potential ${\cal U}$ has a $\phi$ dependence only through the cubic term 
$R^3\cos{3\phi}$, that is, the $\Phi^3+{\Phi^*}^3$ term. 
The fact $-b<0$ means that ${\cal U}$ has a minimum at $\phi=0$ 
mod $2\pi/3$, if $R$ is not zero. 
Now we consider the case of $\phi=0$. 
Figure~\ref{pure-gauge-U} show the $R$ dependence of ${\cal U}/\Lambda^4$ at 
$\phi=0$. 
Three cases of $T/T_0=0.99, 1, 1.01$ are shown by solid curves 
from the top to the bottom, respectively, where $T_0=270$~MeV. 
For each $T$, ${\cal U}$ has two local minima. 
When $T<T_0$ a global minimum is always located at $R=0$, but 
at $T=T_0$ the location jumps from $R=0$ to 0.45. 
Thus, a first-order deconfinement transition takes place at $T=T_0$. 
Due to the $R^3\cos{3\phi}$ term, at $T>T_0$, 
there are three $\mathbb{Z}_3$ global minima at 
$(R,\phi)=(R_0,0~{\rm mod}~2\pi/3)$, 
where $R_0$ is a value between 0.45 and 1. 
Hence, the ground state of the pure gauge system 
has a 3-fold degeneracy above $T_0$, but 
no degeneracy below $T_0$. 
In the pure gauge system, thus, 
the high-$T$ phase is distinguishable from the low-$T$ one
by the number of the vacuum degeneracy. 

Next, we consider the system with dynamical quarks. 
As a feature of the system, $\Omega(\theta)$  
is a smooth function of $\theta$ below $T_{\rm E}$, 
as shown in Fig.~\ref{theta_Omega_stable}. 
In order to understand this property, we consider the case of small $T$.
First, we assume that  
$\Phi$ tends to zero 
and $\sigma$ does to a finite value 
$\sigma_0$ in the limit of small $T$. 
This is a natural assumption and justified below. 
In the case of small but nonzero $T$, 
the stationary conditions \eqref{eq:SC} for $\Phi$ and 
$\sigma$ have $T$ dependence through 
factors $e^{-\beta E({\bf p})}$ and $e^{-\beta a}$. 
The factor $e^{-\beta E({\bf p})}$ has a maximum $e^{-\beta M}$ 
at ${\bf p}=0$, and near $T=0$ the maximum is very close to 
$\epsilon \equiv e^{-\beta M_0}$ with 
$M_0=m_0-2G_{\rm s}\sigma_0$. 
Hence, the ${\bf p}$ integration of $e^{-\beta E({\bf p})}$ is of order 
$\epsilon$. 
Further, the factor $e^{-\beta a}$ is of order $\epsilon^2$ 
because of $a \approx 2 M_0$. It is then natural to expand 
$\Phi$ and $\sigma$ by powers of $\epsilon$: namely,  
$\Phi= \Phi_1 \epsilon +\Phi_2 \epsilon^2 \cdots$ and 
$ \sigma = \sigma_0 + \sigma_1 \epsilon \cdots$.  
One can then derive equations for 
coefficients $\Phi_i$ and $\sigma_i$ of the $\epsilon$ expansion, 
order by order, from the stationary conditions \eqref{eq:SC}. 
The equations for the leading-order solutions, $\sigma_0$ and $\Phi_1$, are 
\begin{eqnarray}
\sigma_0 &=&- {6N_f}\int \frac{d^3{\rm p}}{(2\pi)^3}
{M_0 \over {E_0({\bf p})}} , 
\label{Ernew1}
\\ 
\Phi_1 &=& \left\{ \frac{N_f}{b} \int \frac{d^3{\rm p}}{(2\pi)^3}
e^{-\beta E_0({\bf p})+\beta M_0}\right\} e^{-i\theta } ,  
\label{Er1}
\end{eqnarray}
where $E_0({\bf p})=\sqrt{{\bf p}^2+M_0^2}$. 
Equation \eqref{Ernew1} does not include $T$ and $\theta$, 
so that the solution $\sigma_0$ is a constant, as expected. 
In Eq.~\eqref{Er1}, $\Phi_1 \epsilon$ tends to zero 
as $T$ decreases, as expected. 
The absolute value $|\Phi_1|$ does not depend on 
$\theta$, while the phase of $\Phi_1$ does. 
Inserting the leading-order solutions into 
the stationary conditions \eqref{eq:SC}, we can get 
higher-order solutions; for example, $\sigma_1=0$ and 
\begin{eqnarray}
\Phi_2  &=& \left\{ 
2 |\Phi_1|^2 + { N_f \over b } 
\int \frac{d^3{\rm p}}{(2\pi)^3} e^{-2\beta E_0({\bf p})+2\beta M_0} 
\right\}  e^{2i\theta },
\label{phi2}
\\
\sigma_2 &=& { 12N_f |\Phi_1|  \over {C_0} } 
\int \frac{d^3{\rm p}}{(2\pi)^3}{ M_0 \over {E_0({\bf p})} }
e^{-\beta E_0({\bf p})+\beta M_0} ,
\label{Er3}
\\
\sigma_3 &=& {12 N_f  \over {C_0} } 
\int \frac{d^3{\rm p}}{(2\pi)^3}{ M_0 \over {E_0({\bf p})} }
\cos(3\theta) 
\notag \\
&& \times \left\{
|\Phi_2|e^{-\beta E_0({\bf p})+\beta M_0} 
+2|\Phi_1|e^{-2\beta E_0({\bf p})+2\beta M_0} 
+e^{-3\beta E_0({\bf p})+3\beta M_0} 
\right\}
\label{Er4}
\end{eqnarray}
with 
\begin{eqnarray}
C_0&\equiv &1-{12N_f}G_{\rm s}\int \frac{d^3{\rm p}}{(2\pi)^3}
{{\bf p}^2\over{(E_0({\bf p}))^3}}\sim 0.4, 
\label{Erv1}
\end{eqnarray}
where $\Phi_3$ has not been exhibited, because it does not contribute to 
$\Omega$ up to $\epsilon^3$, as shown below. 
The second-order solutions, 
$|\Phi_2|$ and $\sigma_2$, do not depend on $\theta$, 
but the phase of $\Phi_2$ and the third-order solution $\sigma_3$ do. 
Thus, unique set of solutions, 
$\sigma = \sum_{n=0} \sigma_n \epsilon^n $ 
and $ \Phi = \sum_{n=1} \Phi_n \epsilon^n $, is obtained at small $T$. 
The thermodynamic potential $\Omega$ is given by inserting the set of 
solutions into Eq.~\eqref{eq:E12}. 
Hence, the potential, $\Omega = \sum_{n=0} \Omega^{(n)}\epsilon^n$, 
thus obtained is unique and then smooth in $\theta$; 
note that $|F| \ll 1$ and $|G| \ll 1$ and then 
$\log(1+F) \approx F$ and $\log(1+G) \approx G$ have no singularity. 
The factors $F$ and $G$ can also be expanded into 
$F=\sum_{n=2} F_n\epsilon^n$ and $G=\sum_{n=2} G_n\epsilon^n$ with  
\begin{eqnarray}
F_2 &=&  3|\Phi_1| e^{-\beta E_0({\bf p})+\beta M_0}, ~~~
G_2=      -6|\Phi_1|^2,
\label{Er7} \\ 
F_3 &=& \left\{ 3|\Phi_2| e^{-\beta E_0({\bf p})+\beta M_0} 
       + 3|\Phi_1| e^{-2\beta E_0({\bf p})+2\beta M_0}
       + e^{-3\beta E_0({\bf p})+3\beta M_0}
       \right\}  e^{3i\theta} ,	
\label{Er8}\\
G_3 &=& - 4  
(3 |\Phi_1||\Phi_2| - 2|\Phi_1|^3) 
\cos(3\theta), \cdots. 
\label{Er9}
\end{eqnarray}
Eventually, the $\Omega^{(n)}$ are given as 
\begin{eqnarray}
\Omega^{(0)} &=&G_{\rm s} \sigma_0^2 -6N_f \int \frac{d^3{\rm p}}{(2\pi)^3}
           E_0 ({\bf p}),~~~
\Omega^{(1)}=0, 
\label{Er5} \\
\Omega^{(2)} &=&2 G_{\rm s} \sigma_0\sigma_2
+12G_{\rm s} N_f \sigma_2 
\int \frac{d^3{\rm p}}{(2\pi)^3}{ M_0 \over {E_0({\bf p})} } 
-4 N_f T 
\int \frac{d^3{\rm p}}{(2\pi)^3}F_2 - b T G_2, 
\label{Er6} \\
\Omega^{(3)} &=&2 G_{\rm s} \sigma_0\sigma_3
+12G_{\rm s} N_f \sigma_3
\int \frac{d^3{\rm p}}{(2\pi)^3}{ M_0 \over {E_0({\bf p})} } 
-4 N_f T  \cos(3\theta) 
\int \frac{d^3{\rm p}}{(2\pi)^3}|F_3|
\notag \\
&& - b T G_3, 
\label{Er7} 
\end{eqnarray}
up to order $\epsilon^3$. 
Thus, $\Omega$ has no $\theta$ dependence up to order $\epsilon^2$, and 
the $\theta$-dependence appears first at order $\epsilon^3$ through 
the factor $\cos(3\theta)$. 
Therefore, $\Omega$ is a smooth function of $\cos(3\theta)$ at small $T$, 
but the $\theta$ dependence is very weak, 
as explicitly shown in Fig.~\ref{theta_Omega_stable}.

Above $T_{\rm E}$, $\Omega$ has a cusp at $\theta =\pi/3$ (mod $2\pi/3$), 
as shown in Fig. \ref{theta_Omega_stable}. 
To understand the nature of the discontinuity, 
we consider the case of high $T$, extend $N_{f}$ 
to be a continuous variable and take 
the small $N_{f}$ limit. 
Since $\Omega^{\rm vac}$ and $\Omega^{\rm ther}$ 
are proportional to $N_f$, it can be treated perturbatively. 
To the zeroth order of $N_{f}$, the thermodynamic potential is reduced to 
that for the pure gauge theory. 
In this gauge dominant case, the thermodynamic potential has a global minimum 
at $R=R_0$ satisfying $0.45 < R_0 <1$, 
as shown in Fig.~\ref{pure-gauge-U}.  
Therefore, there exit three unperturbed solutions 
$\Phi_{k}\equiv |\Phi_{k}| e^{i \phi_k}= R_0e^{i 2 \pi k/3}~(k=0, \pm 1)$, 
since ${\cal U}$ is invariant under the ${\mathbb Z}_3$ transformation.

Inserting the zeroth solutions $\Phi_k = R_0e^{i 2 \pi k/3}$ ($k=0, \pm 1$) 
into $\Omega$ of Eq. (\ref{eq:K3}), one can get three kinds of thermodynamic 
potentials $\Omega_{k}(\theta)$ ($k=0, \pm 1$). 
For simplicity, we consider the limit of $R_0\to 1$. 
The thermal fermionic parts $\Omega_{k}^{\rm ther}(\theta)$ are 
\begin{align}
\Omega_{k}^{\rm ther}(\theta,\sigma ) 
\sim - & \frac{6 N_f}{\beta} \int \frac{d^3{\rm p}}{(2\pi)^3}
         \Bigl[ 
          \ln~ [1 +  e^{-\beta E({\bf p})}e^{i\theta+i2\pi k/3}]
          \notag\\
         &+
          \ln~ [1 +  e^{-\beta E({\bf p})}e^{-i\theta-i2\pi k/3}]
	      \Bigl] , 
\label{eq:omega-log-2b} 
\end{align}
where $E({\bf p})=\sqrt{{\bf p}^2+(m_0-2G_{\rm s}\sigma)^2}$.  
The chiral condensate $\sigma$ in Eq. (\ref{eq:omega-log-2b}) is determined 
by the stationary condition $\partial \Omega/\partial \sigma=0$:
\begin{align}
\sigma&=-2N_f \int \frac{d^3{\rm p}}{(2\pi)^3}\frac{3 M }{E ({\rm p})}
D({\bf p},T,\theta) 
\label{con-sigma}
\end{align}
with
\begin{align}
D({\bf p},T,\theta)&=1-\frac{1}{e^{\beta E({\bf p})}e^{-i\theta-i2\pi k/3}+1}
    -\frac{1}{e^{\beta E({\bf p})}e^{i\theta+i2\pi k/3}+1}.
\label{D}
\end{align}
Equations \eqref{con-sigma} and \eqref{D} show that $\sigma$ is small 
at high $T$, independently of $N_{f}$, 
because the factor $D({\bf p},T,\theta)$ tends to zero in the limit of high $T$. 
In the numerical calculations for $N_f=2$, 
$\Phi_k$ tends to $e^{i2\pi k/3}$ and $\sigma$ becomes small as $T$ increases. 
In this sense the present analysis is consistent with the numerical results.

\begin{figure}[htbp]
\begin{center}
\includegraphics[width=0.3\textwidth,angle=-90]{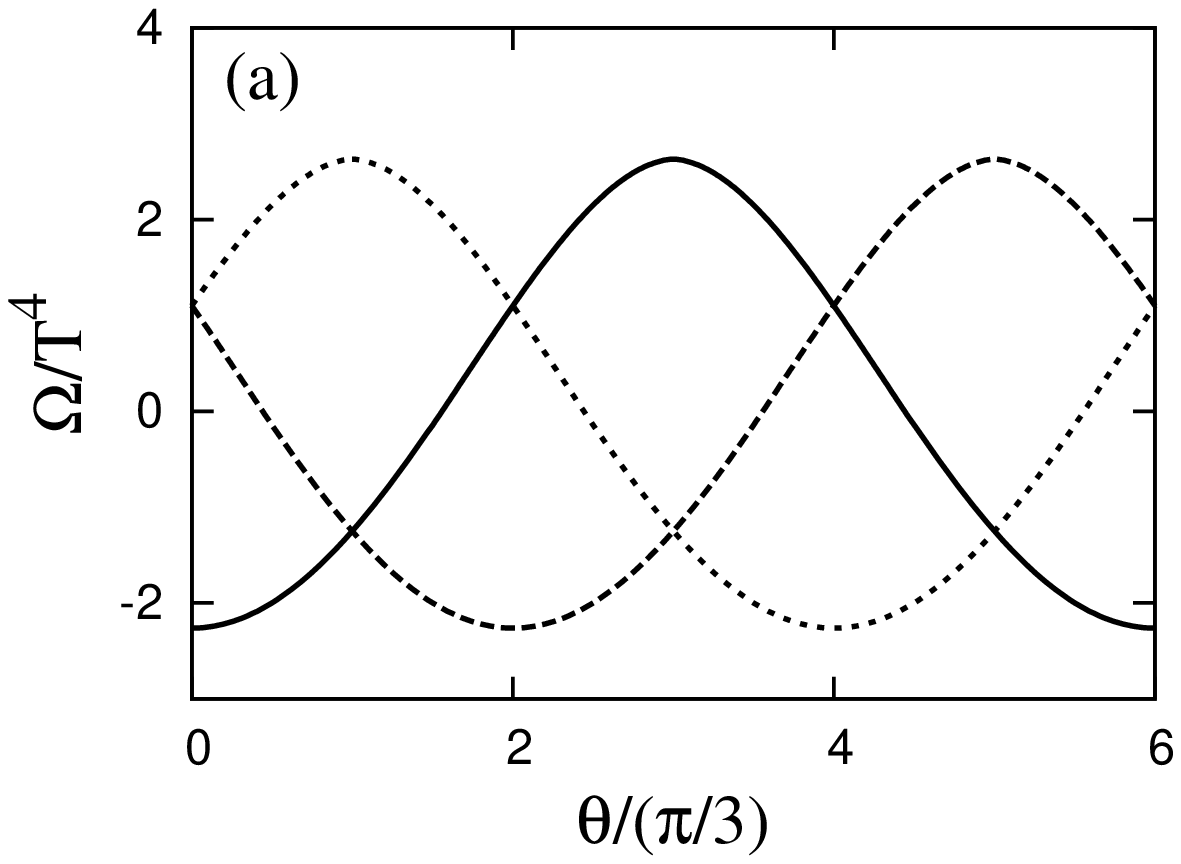}
\includegraphics[width=0.3\textwidth,angle=-90]{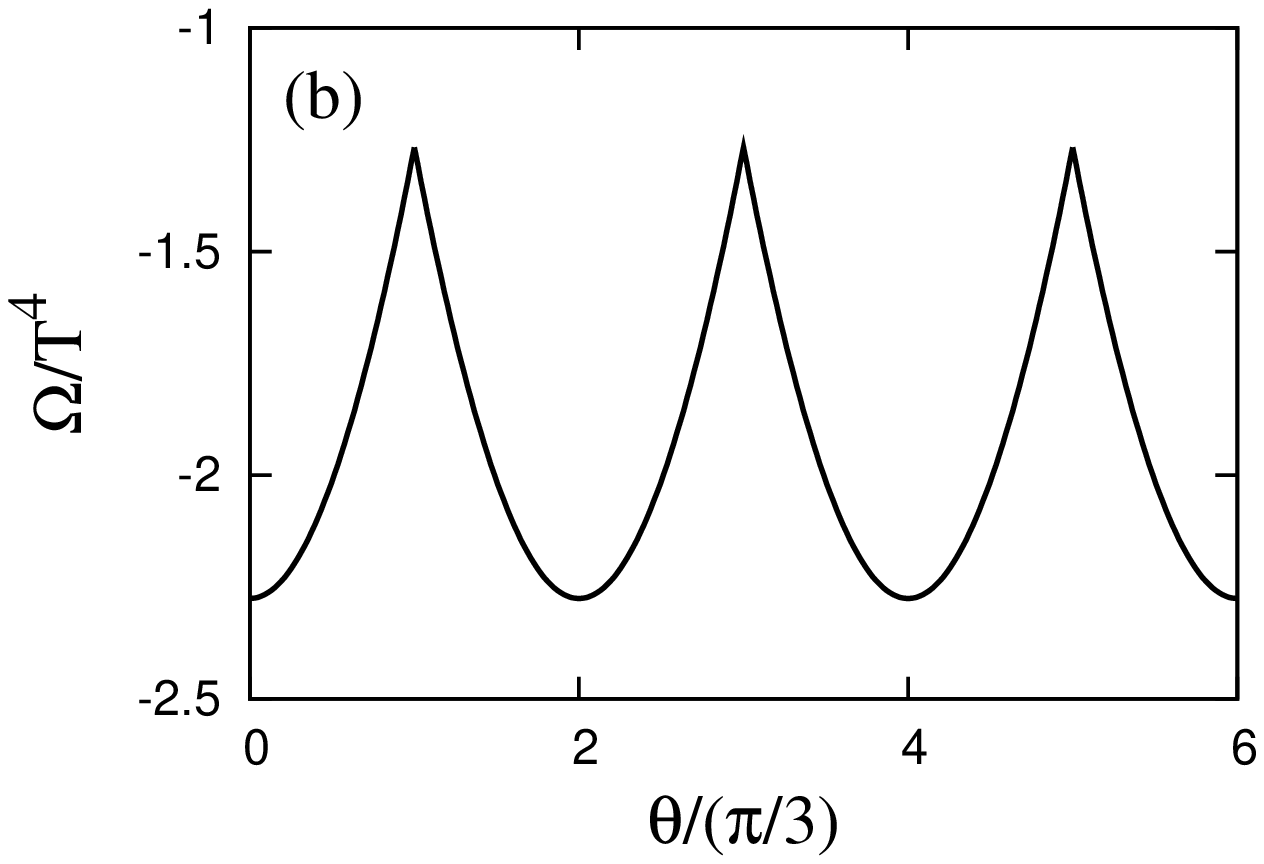}
\end{center}
\caption{Perturbative solutions to 
the thermodynamic potential in its thermal part: 
(a) the smooth solutions $\Omega_k^{\rm ther}$ ($k=0, \pm 1$) 
and (b) the ground-state solution $\Omega_{\rm GS}^{\rm ther}$. 
In (a), the solid, the dotted and the dashed curves show the cases of $k=0$, $k=+1$ and $k=-1$, respectively. 
 }.
\label{perturbative}
\end{figure}

The solutions $\Omega_{k}(\theta)$ 
are smooth, but each solution is periodic only with period $2 \pi$. 
Thus, each of the solutions does not have the RW periodicity. 
This does not mean that a set of the three solutions does not keep the RW periodicity. 
In order to understand this clearly, we consider the vanishing quark mass limit of $M=0$ where the analytic forms of $\Omega_{k}^{\rm ther}(\theta)$ 
are available:
\begin{align}
\Omega_{k}^{\rm ther}(\theta) = \frac{6 N_f}{\beta^4 \pi^2} 
                \Bigl[ {\rm Li}_4(e^{i(\theta-2\pi k/3)}) 
                + {\rm Li}_4(e^{-i(\theta-2\pi k/3)}) 
                \Bigl],  
\label{eq:omega-log-5} 
\end{align}
where ${\rm Li}_4(z)$ is the polynomial logarithm defined by 
${\rm Li}_4(z)=\sum_{n=1}^{\infty} z^n/n^4$. 
As shown in Fig. \ref{perturbative} (a), the three solutions 
$\Omega_{k}^{\rm ther}(\theta)$ 
are smooth but periodic with 
period $2\pi$, and $\Omega_{\pm 1}^{\rm ther}(\theta)$ 
are obviously ${\mathbb Z}_3$ images 
of $\Omega_{0}^{\rm ther}(\theta)$. 
As shown in Fig. \ref{perturbative} (b), 
if the lowest solution is taken at each $\theta$, 
the ground state (GS) thus connected, $\Omega_{\rm GS}^{\rm ther}(\theta)$, 
is periodic with period $2\pi/3$ but not smooth at 
$\theta=\pi/3$ (mod $2\pi/3$). 
As a result of this mechanism, 
$\Omega_{\rm GS}(\theta)\equiv \Omega^{\rm vac}+\Omega_{\rm GS}^{\rm ther}(\theta)+{\cal U}$ is not smooth at $\theta=\pi/3$ (mod $2\pi/3$), that is, 
the RW phase transition appears there. 
We call this mechanism the RW mechanism. 
In the high $T$ case, thus, 
the ground state preserves the RW periodicity, i.e. 
the extended ${\mathbb Z}_3$ symmetry, although 
each $\Omega_{k}^{\rm ther}(\theta)$ does not. 
Eventually, the extended ${\mathbb Z}_3$ symmetry is held at any temperature.

The analyses above based on perturbation are essentially the same as 
Roberge and Weiss did with 
perturbative QCD within its one-loop approximation~\cite{RW,Weiss}. 
Actually, the one-loop solutions agree with 
$\Omega_{k}^{\rm ther}(\theta)$ ($k=0, \pm 1$) of Eq. \eqref{eq:omega-log-5}. 
Thus, at high temperature, the system has three 
${\mathbb Z}_{3}$ vacua, and they appears alternatively as $\theta$ varies 
from 0 to $\pi$. 
This is the RW mechanism, and when $\theta=0$ 
the system belongs to one of the ${\mathbb Z}_{3}$ vacua.

\begin{figure}[htbp]
\begin{center}
 \includegraphics[width=0.3\textwidth,angle=-90]{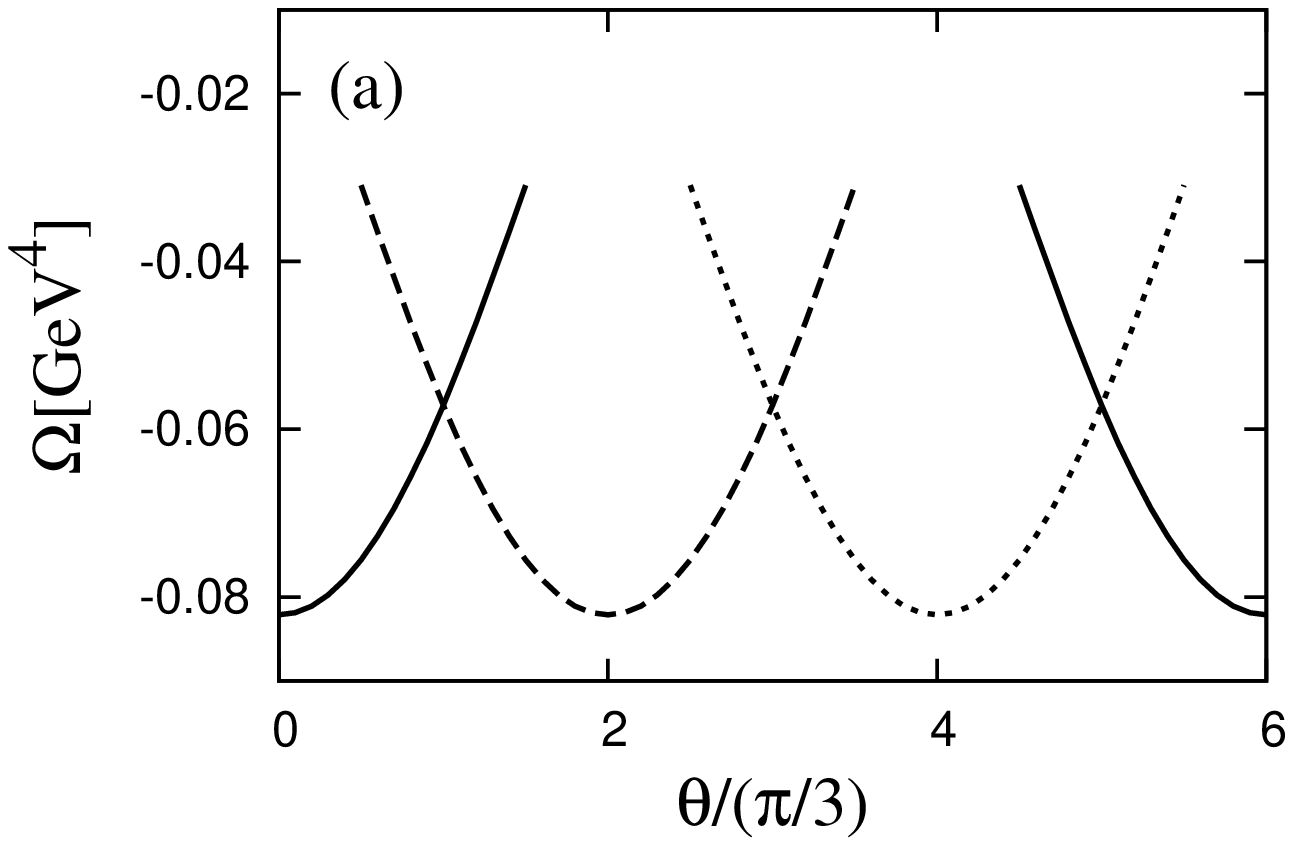}
 \includegraphics[width=0.3\textwidth,angle=-90]{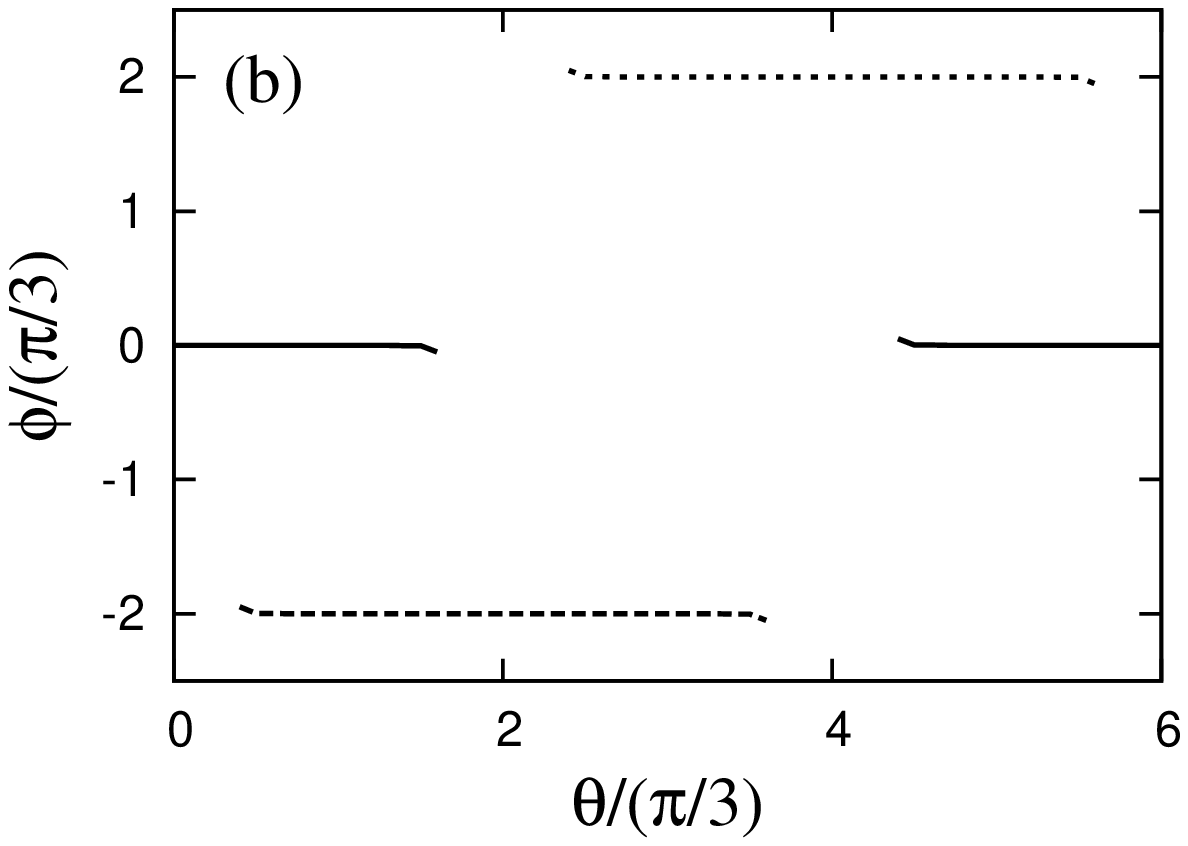}
\end{center}
\caption{$\theta$ dependence of solutions to the stationary conditions 
in the case of $T=400$~MeV: 
(a) represents $\Omega_{k}$ and (b) does $\phi_k$. 
The solid, the dotted and the dashed curves show the cases of $k=0$, $k=+1$ and $k=-1$, respectively. 
}
\label{theta_Omega_unstable}
\end{figure}

The RW mechanism appears also 
in the full PNJL calculation free from perturbation. 
However, the situation is more complicated as shown below. 
The stationary conditions \eqref{eq:SC} give three sets of 
solutions, $\sigma_k$ and $\Phi_k=|\Phi_k|e^{i \phi_k}$ ($k=0, \pm 1$), 
as expected.  
Inserting the solutions into Eq.~\eqref{eq:K3}, we have 
three solutions $\Omega_{k}$. 
Figure~\ref{theta_Omega_unstable} shows 
$\theta$-dependence of $\Omega_k$ and $\phi_k$ in the case of 
$T=400$~MeV. Surely, there exist three kinds of solutions. 
At each $\theta$, however, at least one of the three disappears; 
the solution $\Omega_0$ vanishes in the region 
of $0.53\pi < \theta < 1.46\pi$, and  
the region is shifted by either $2\pi/3$ or $-2\pi/3$ for other solutions. 
Whenever $\Omega_k$ exists, $\phi_k$ is about $2\pi k/3$ 
almost independently of $\theta$, while $|\Phi_k|$ depends on $\theta$. 
The ground state $\Omega_{\rm GS}$ is composed of $\Omega_{0}$ in region (I) 
$-\pi/3 \le \theta \le \pi/3$ (mod $2\pi$), $\Omega_{-1}$ in region 
(II) $\pi/3 \le \theta \le \pi$ (mod $2\pi$), and $\Omega_{1}$ in region 
(III) $-\pi \le \theta \le -\pi/3$ (mod $2\pi$).

The RW mechanism are analogous 
to the Dashen phenomenon~\cite{Dashen} 
in the so-called $\Theta$-vacuum. 
Following Witten's analysis~\cite{Witten} on the Dashen phenomenon, 
we can discuss the spontaneous breaking of the $C$ symmetry 
in the RW mechanism. 
The $C$ transformation changes the sign of $\theta$. 
Hence, for the thermodynamic potential $\Omega(\theta)$ with $\theta$ fixed, 
$C$ is a symmetry of $\Omega(\theta)$ only at $\theta=0$ or $\theta=\pi$;
note that $\theta=\pi/3$ mod $2\pi /3$ has the same property as 
$\theta=\pi$ because of the RW periodicity. 
If two of the solutions $\Omega_k(\theta)$ 
cross each other at $\theta=0$ and $\theta=\pi$, 
each solution is $C$-violating and 
$C$ interchanges the two solutions there. 
This Witten's argument on $C$ violation can be 
explicitly confirmed in this case, 
as mentioned below. 
The charge conjugation $C$ transforms 
the sign of $\psi$, and 
hence $\psi=0$ is invariant under $C$. In addition, 
$\psi=\pi$ is also invariant under $C$ because 
$\psi=\pi$ is identical with $\psi=-\pi$. 
The solution $\Omega_{0}(\theta)$ is $C$-conserving, 
since it has $\psi=0$ at $\theta=0$. 
Meanwhile, $\Omega_{\pm 1}(\theta)$ are $C$-violating solutions, because 
in these solutions $\psi$ is neither 0 nor $\pi$ at $\theta=\pi$. 
As shown in Fig.~\ref{theta_Omega_unstable}(a), 
the $C$-violating solutions $\Omega_{\pm 1}(\theta)$ cross each other 
at $\theta=\pi$, while the $C$ conserving solution 
$\Omega_0(\theta)$ has no crossing at $\theta=0$. 
Thus, the $C$ symmetry is conserved at $\theta =0$, but 
spontaneously broken at $\theta =\pi$. 
This $C$ symmetry breaking appears also at $\theta =\pm \pi/3$ as 
a consequence of the RW periodicity. 

When $T<T_{\rm E}$, the $\theta$-odd quantity 
$\psi$ is zero at $\theta =0$ and $\pi$, 
because $\psi$ is a smooth function of $\theta$ satisfying Eq.~\eqref{odd}. 
Hence, the $C$-symmetry is preserved there. 
Furthermore, $\psi =0$ at $\theta = k\pi /3$ 
with integer $k$ as a result of the RW periodicity. 
As seen in Sec. \ref{sec:RW-phase-transition}, 
$\theta$-dependence of $\Omega(\theta)$ is weak, and then 
$\psi\sim 0$. Therefore, $\phi =\psi -\theta \sim -\theta$ at low $T$. 
On the contrary, when $T>T_{\rm E}$, $\phi$ is almost constant 
in each of regions (I), (II), (III). 
The $T>T_{\rm E}$ regime is thus 
distinguishable from the $T<T_{\rm E}$ one by 
the $\theta$-dependence of $\phi$ (or $\psi$).

\subsection{Order of RW phase transition at endpoint}
\label{sec:order}

Susceptibilities $\chi_{ij}$ of $\sigma$, $R=|\Psi |$ and $\psi ={\rm arg}(\Psi)$ can be written as~\cite{Fukushima,Sasaki,Kashiwa1} 
\begin{eqnarray}
\chi_{ij} &=& (K^{-1})_{ij}
~~~~(i,j=\sigma, R, \psi),
\end{eqnarray}
where  
\begin{eqnarray}
K = \left(
\begin{array}{cccc} 
\frac{\partial^2 \Omega}{\partial \sigma^2} 
& 
\frac{\partial^2 \Omega}{\partial \sigma \partial R} 
& 
\frac{\partial^2 \Omega}{\partial \sigma \partial \psi} \\

\frac{\partial^2 \Omega}{\partial R \partial \sigma} 
& 
\frac{\partial^2 \Omega}{\partial R^2} 
& 
\frac{\partial^2 \Omega}{\partial R \partial \psi} \\
\frac{\partial^2 \Omega}{\partial \psi \partial \sigma} 
& 
\frac{\partial^2 \Omega}{\partial \psi \partial R} 
& 
\frac{\partial^2 \Omega}{\partial \psi^2} \\
\end{array}
\right) ,
\end{eqnarray}
is a symmetric matrix of curvatures of $\Omega$ and $K_{ij}$ is an 
$(i,j)$ element of curvature matrix $K$. 
Matrix elements $K_{\sigma \psi}$ and $K_{R\psi}$ are $\theta$-odd and then 
zero at $T\leq T_{\rm E}$ and $\theta=\pi/3~({\rm mod}~2\pi/3)$, as shown 
in Eq.~\eqref{odd}. 
Hence, the $\chi_{ij}$ is obtained there as 
\begin{eqnarray}
\chi_{\psi\psi}=\frac{1}{K_{\psi\psi}}, \quad 
\chi_{ij} = (K_2^{-1})_{ij}~~~(i,j=\sigma, R),
\label{sus-2}
\end{eqnarray}
where 
\begin{eqnarray}
K_2 = \left(
\begin{array}{cccc} 
\frac{\partial^2 \Omega}{\partial \sigma^2} 
& 
\frac{\partial^2 \Omega}{\partial \sigma \partial R} 
\\
\frac{\partial^2 \Omega}{\partial R \partial \sigma} 
& 
\frac{\partial^2 \Omega}{\partial R^2} 
\end{array}
\right).
\end{eqnarray}
At the critical point 
$(\theta, T)=(\pi /3~{\rm mod}~2\pi/3,T_{\rm E})$, thus, 
the susceptibilities of $\sigma$ and $R$ decouple from that of $\psi$, the order parameter of the phase transition.

\begin{figure}[htbp]
\begin{center}
 \includegraphics[width=0.4\textwidth]{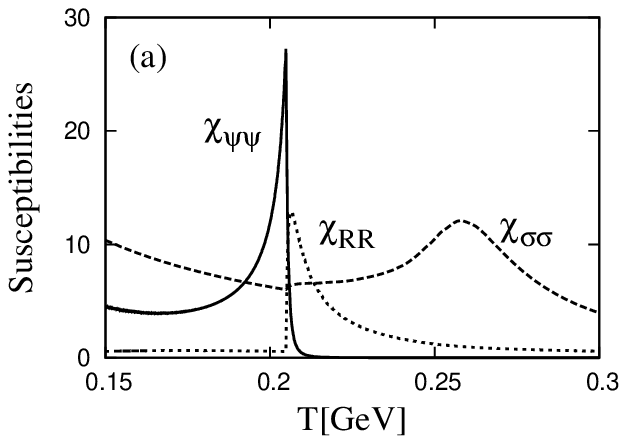}
 \includegraphics[width=0.4\textwidth]{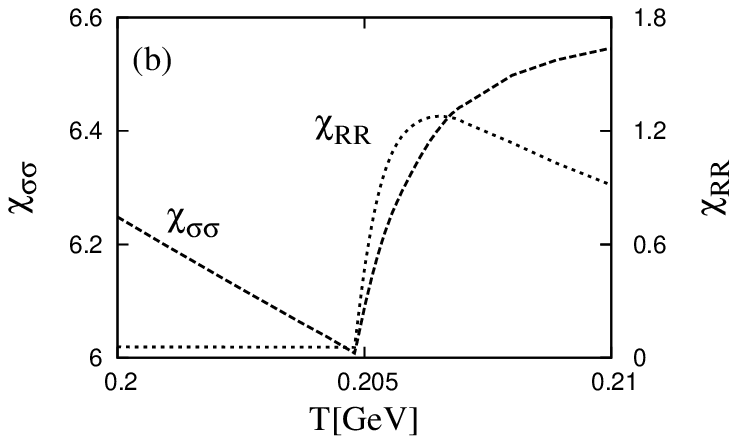}
 \includegraphics[width=0.4\textwidth]{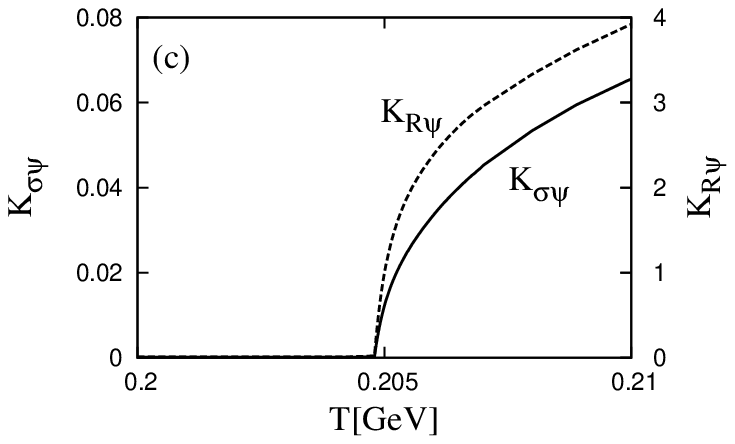}
\end{center}
\caption{ 
(a) Susceptibilities of the chiral condensate $\sigma$ (dashed curve), 
the absolute value $R$ of the modified Polyakov loop (dotted curve) 
 and the phase of the modified Polyakov loop (solid curve) 
as a function of $T$ at $\theta =\pi$. 
$\chi_{\sigma\sigma}$ is divided by $T^2$, while $\chi_{RR}$ and $\chi_{\psi\psi}$ are multiplied by $10T^4$ and $0.2T^4$, respectively. 
(b)  Susceptibilities of the chiral condensate $\sigma$ (dashed curve) and 
the absolute value $R$ of the modified Polyakov loop (dotted curve) 
near the critical temperature $T_{\rm E}$. 
$\chi_{\sigma\sigma}$ is divided by $T^2$, while $\chi_{RR}$ and $\chi_{\psi\psi}$ are multiplied by $T^4$. 
(c) Off-diagonal elements $K_{\sigma\psi}$ (solid curve) 
and $K_{R\psi}$ (dashed curve) are shown as functions of $T$. 
$K_{\sigma\psi}$ and $K_{R\psi}$ are divided by $T$ and $T^4$, respectively. }
\label{sus}
\end{figure}

Figures~\ref{sus}(a) and (b) present $T$ dependence of 
susceptibilities, $\chi_{\sigma \sigma}$, $\chi_{RR}$ and $\chi_{\psi\psi}$, 
at $\theta=\pi $. 
The susceptibility $\chi_{\psi\psi}$ has a divergent peak at $T=T_{\rm E}=0.2048$~GeV. 
This indicates that the RW phase transition is of second order at 
the endpoint $(\theta ,T)=(\pi /3~{\rm mod}~2\pi/3,T_{\rm E})$ 
and also that a $\theta$-odd quantity such as $\psi$ is an order parameter of the phase transition. 
This result is consistent with the LQCD result~\cite{FP}. 

As shown in Eq.~\eqref{sus-2}, the divergence comes from the fact that 
$K_{\psi\psi}=0$ at the endpoint, and the susceptibilities 
of the $\theta$-even quantities such as 
$\chi_{\sigma \sigma}$ and $\chi_{RR}$ are irrelevant to $K_{\psi\psi}$.
Therefore, the divergent behavior does not affect 
$\chi_{\sigma \sigma}$ and $\chi_{RR}$.  
Actually, $\chi_{\sigma \sigma}$ and $\chi_{RR}$ 
have no divergent peak there, as shown in Figs.~\ref{sus}(a) and (b). 
There is no a priori reason that the transition temperature $T_{\sigma}$ defined at the peak position of $\chi_{\sigma\sigma}$ coincides with $T_{\rm E}$ at $\theta=\pi/3~({\rm mod}~2\pi/3)$. 
However, it is shown that the vector-type four-quark interaction and the eight-quark interaction make $T_{\sigma}$ closer to $T_{\rm E}$~\cite{Sakai2}. 

This second-order phase transition is accompanied 
by the spontaneous breaking of charge conjugation ($C$) symmetry.
The $C$ symmetry is a $\mathbb{Z}_2$ symmetry, 
since $\Omega$ is $C$-even ($\theta$-even) 
and then an even function of $C$-odd ($\theta$-odd) quantities such as Im$[\Psi]$ and $\psi$. Thus, the $\mathbb{Z}_2$ symmetry 
is spontaneously broken at $T > T_{\rm E}$. 

In the $\mu_{\rm R}$ region, 
the second-order chiral phase transition 
at the critical endpoint is known to be 
accompanied by a spontaneous $\mathbb{Z}_2$ symmetry breaking~\cite{AY}. 
However, as shown in Refs.~\cite{Fujii,Fujii2}, 
the $\mathbb{Z}_2$ symmetry is not exact such as the $C$ symmetry 
at the RW endpoint, and the flat direction of the effective potential at the critical endpoint is not the $\sigma$ direction but a linear combination of the chiral condensate $\sigma$, the entropy density $s$ and the quark number 
density $n$. 
Therefore, off-diagonal elements of the curvature matrix $K$ 
do not vanish, and therefore susceptibilities of $\sigma$, $s$ and $n$ 
diverge simultaneously at the critical endpoint 
where the determinant of $K$ vanishes. 

As seen in Fig.~\ref{sus}(b), 
$\chi_{\sigma\sigma}$ and $\chi_{RR}$ have cusps at $T_{\rm E}$. 
Since $K_{\sigma\psi}$ and $K_{R\psi}$ are $\theta$-odd, as mentioned above, 
they are zero in the region $T\leq T_{\rm E}$ where the C symmetry is preserved, but 
become finite in the region $T> T_{\rm E}$ where the C symmetry is spontaneously broken 
via the second-order phase transition. 
Eventually, as shown in Fig.~\ref{sus}(c), 
$K_{\sigma\psi}$ and $K_{R\psi}$ have cusps at $T_{\rm E}$. 
This means that in principle 
all the susceptibilities have cusps at $T_{\rm E}$, 
because they are given by the inverse of the curvature matrix $K$. 
However, this singular behavior is masked by the divergence 
in $\chi_{\psi\psi}$. 

\begin{figure}[htbp]
\begin{center}
 \includegraphics[width=0.3\textwidth,angle=-90]{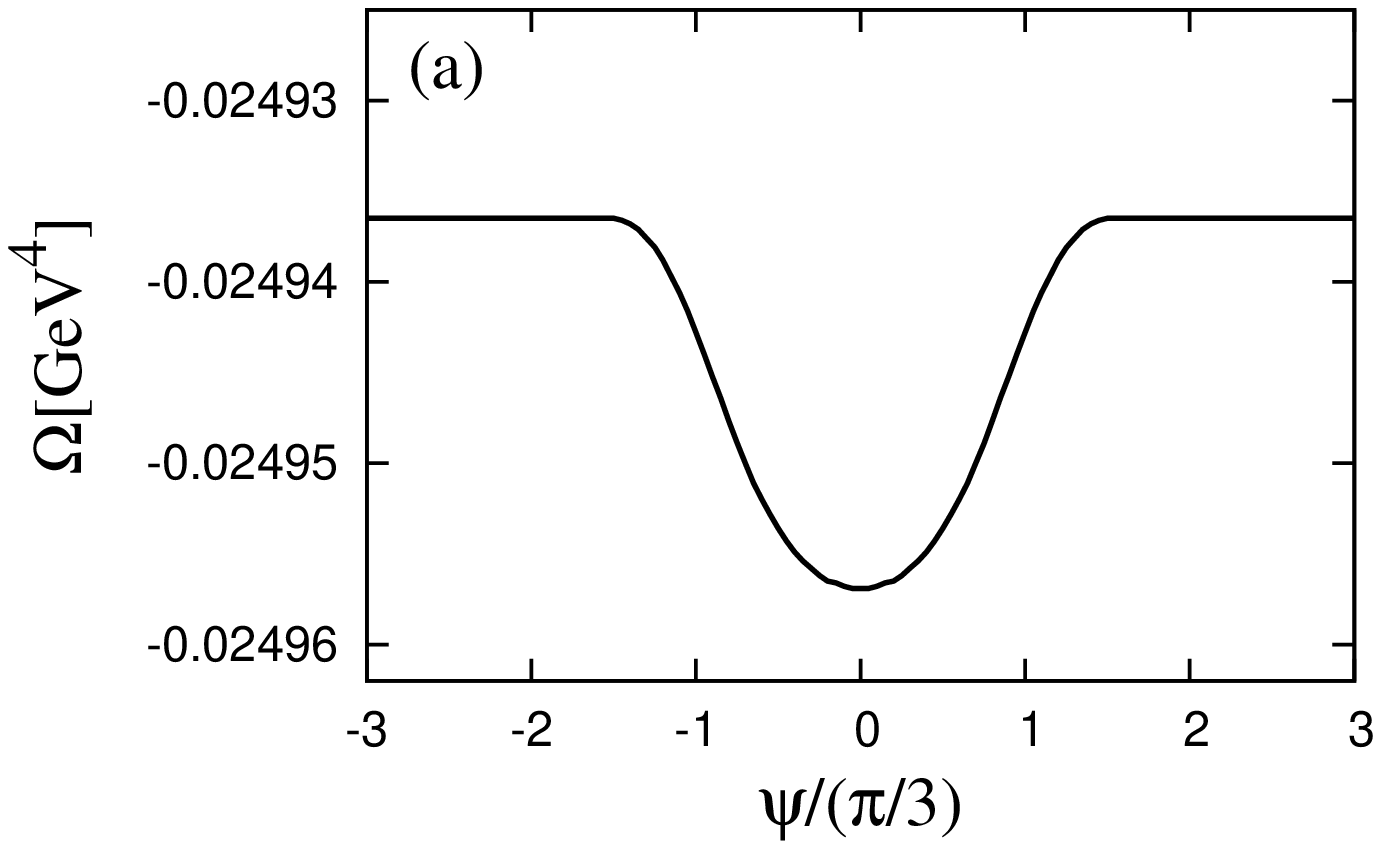}
 \hspace{0.5cm}
 \includegraphics[width=0.3\textwidth,angle=-90]{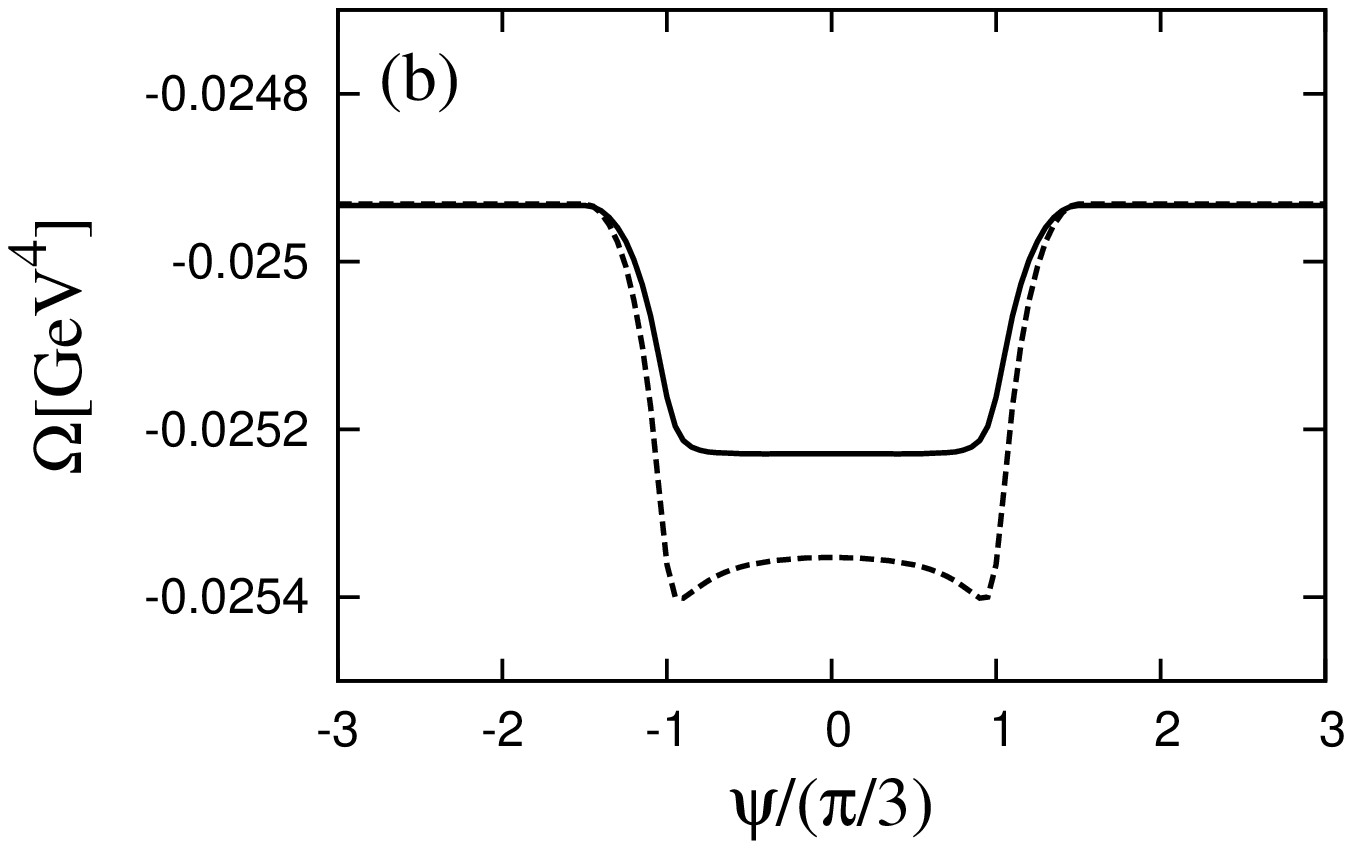}
 \includegraphics[width=0.3\textwidth,angle=-90]{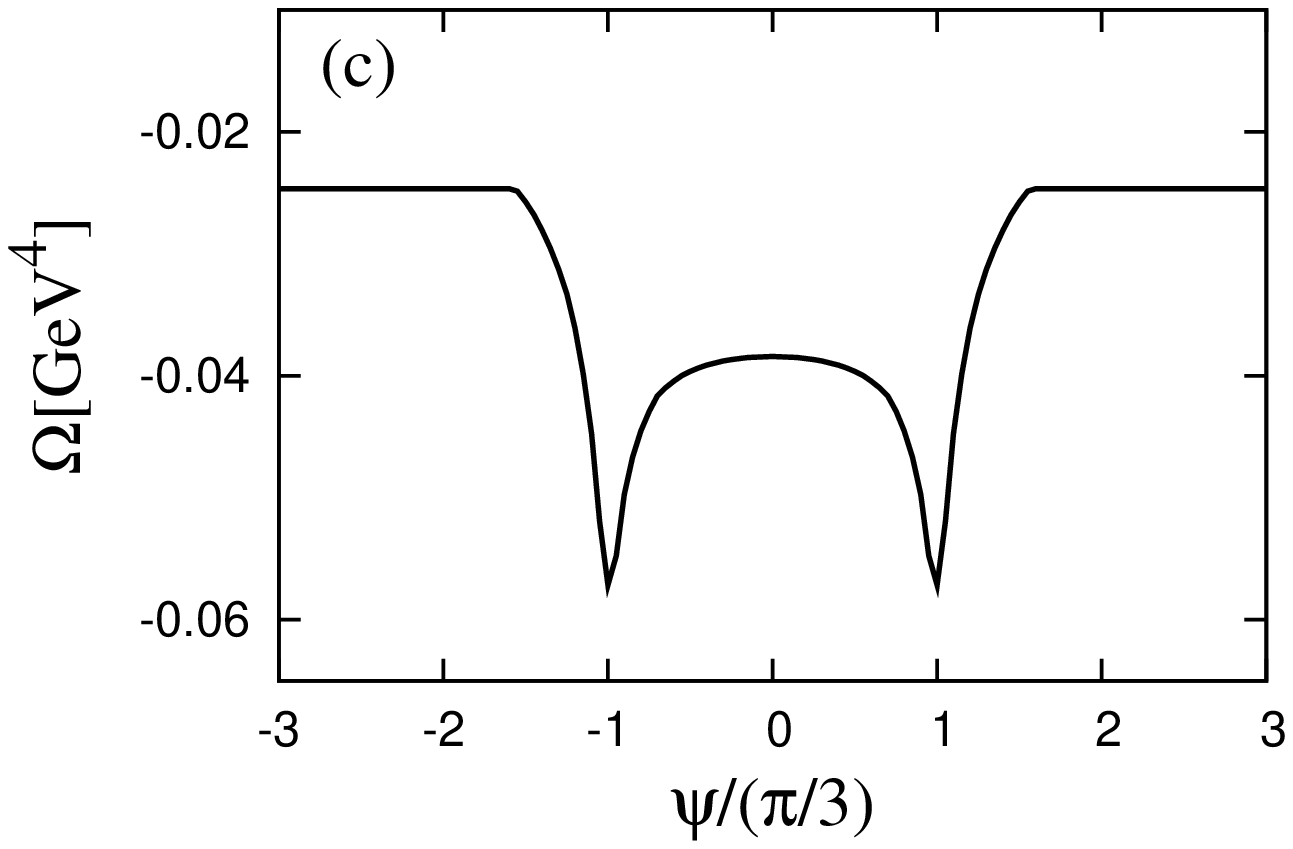}
\end{center}
\caption{Potential surface $\Omega (\psi )$ 
as a function of $\psi$ at $\theta =\pi $:  
(a) represents the case of $T=150$~MeV, 
(b) does two cases of $T=205$ and 215~MeV
and (c) does the case of $T=400$~MeV. 
In (b), the case of $T=205$~MeV ($215$~MeV) is denoted by the solid 
(dashed) curve. 
}
\label{potential_rw}
\end{figure}

The thermodynamic potential $\Omega$ of Eq.~\eqref{eq:K2} is a function 
of variables $R$, $\psi$ and $\sigma$. 
Taking a minimum of $\Omega$ 
in variation of $R$ and $\sigma$ with $\psi$ fixed, 
we can define the potential surface $\Omega(\psi)$ 
as a function of the order parameter $\psi$.  
Figure \ref{potential_rw} shows the potential surface 
at $\theta=\pi$. 
Panels (a)-(c) show the surface in the cases of $T/T_{\rm E}=0.73$, 
$T/T_{\rm E}=1,~1.05$ and $T/T_{\rm E}=1.95$, respectively. 
For the three cases, surely, $\Omega(\psi)$ is $\mathbb{Z}_2$ symmetric under 
the transformation $\psi \to -\psi$. 
In the case of $T/T_{\rm E}=0.73$, there is a minimum at $\psi=0$. 
This minimum can be regarded as a ground state $\Omega_{\rm gr}$. 
The $C$ symmetry is not broken in $\Omega_{\rm gr}$. 
In the case of $T/T_{\rm E}=1.95$, there are two minima at 
$\psi \approx \pm \pi/3$. These correspond to 
solutions $\Omega_{\pm 1}$ 
with $\psi \approx \pm 2\pi/3+\theta \approx \mp \pi /3$ (mod $2\pi$) in 
Fig.~\ref{theta_Omega_unstable}. 
It should be noted that there is no minimum at $\psi\sim \pi$. 
This is consistent with the fact that in Fig.~\ref{theta_Omega_unstable} 
there is no solution with $\phi\sim 0$ at $\theta=\pi$; 
note that $\phi =\psi -\theta$. 
At $T=T_{\rm E}$, the potential surface is flat around $\psi=0$, 
indicating that the RW phase transition is of second order at 
the endpoint $(\theta,T)=(\pi,T_{\rm E})$. 
The same is true also at $\theta =\pi /3~({\rm mod}~2\pi/3)$ as a consequence of the RW periodicity.

\begin{figure}[htbp]
\begin{center}
 \includegraphics[width=0.3\textwidth,angle=-90]{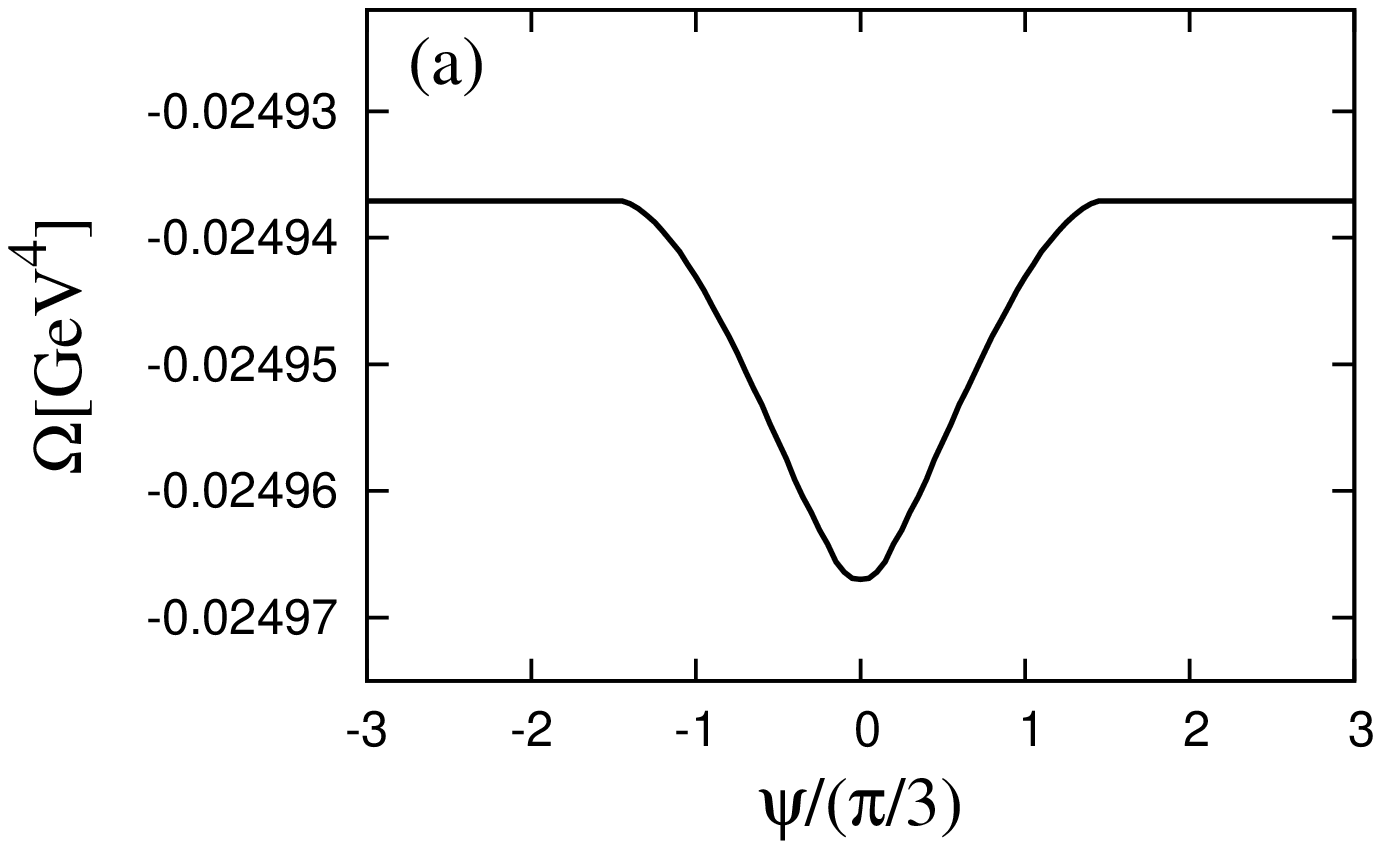}
 \hspace{0.5cm}
 \includegraphics[width=0.3\textwidth,angle=-90]{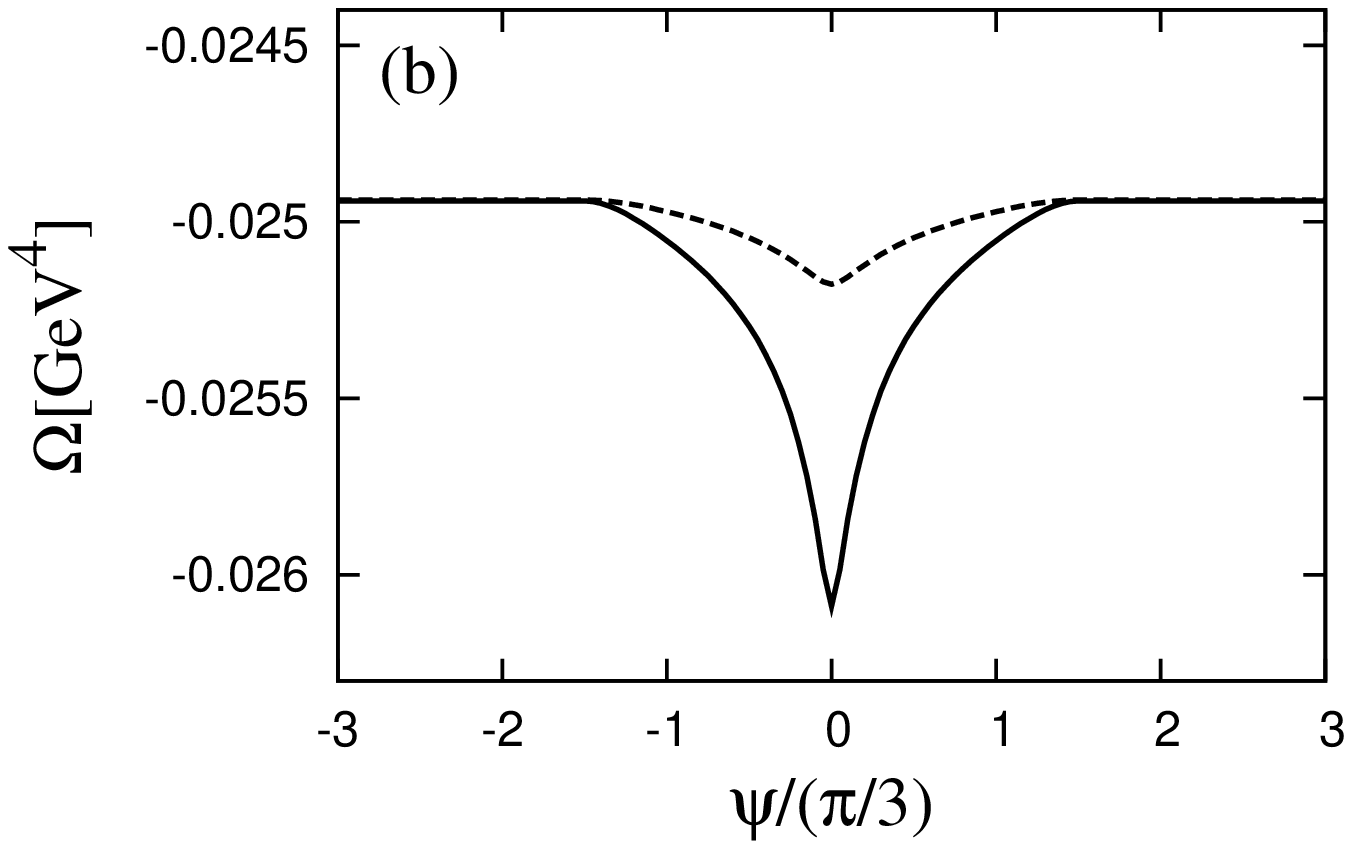}
 \includegraphics[width=0.3\textwidth,angle=-90]{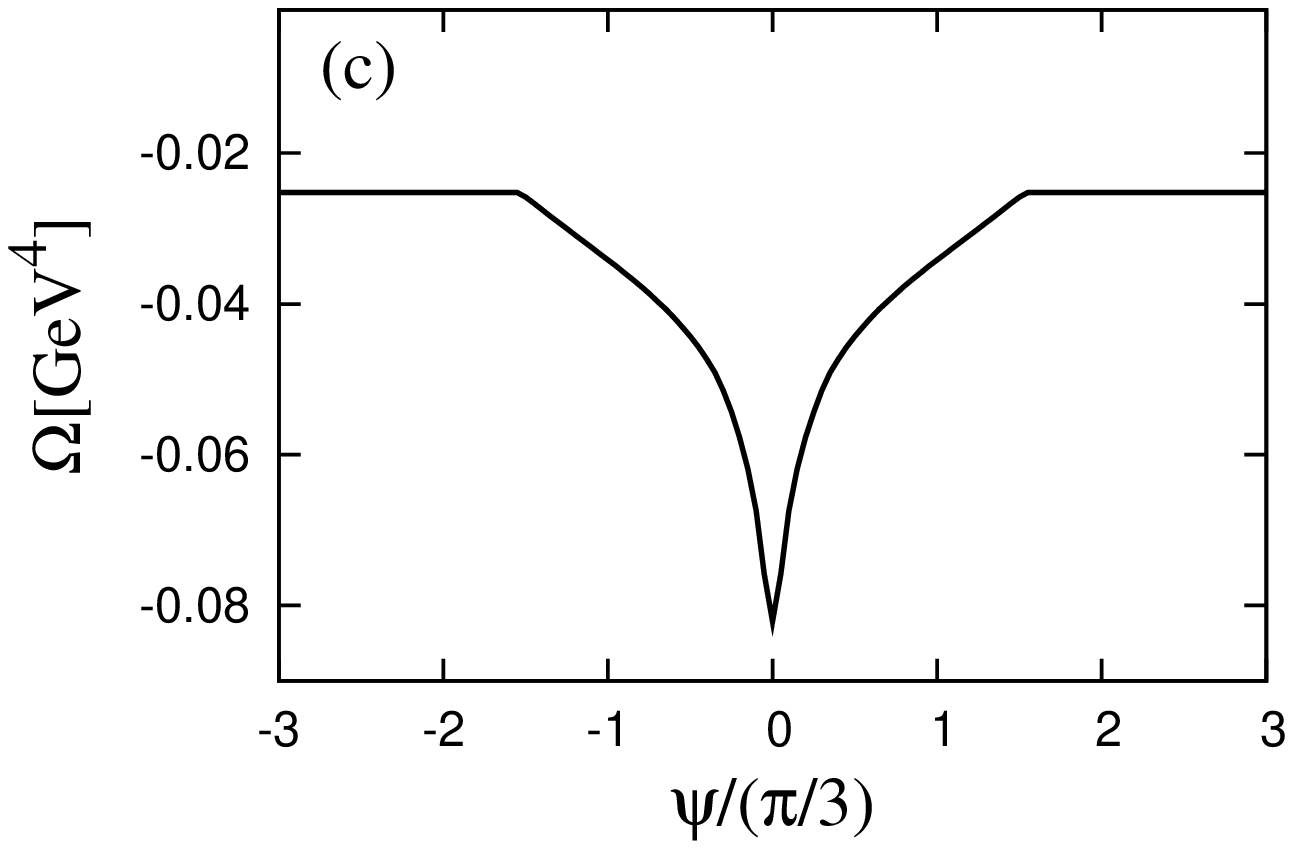}
\end{center}
\caption{Potential surface $\Omega (\psi )$ 
as a function of $\psi$ at $\theta =0$: 
(a) represents the case of $T=150$~MeV, 
(b) does two cases of $T=179$ and 205~MeV 
and (c) does the case of $T=400$~MeV. 
In (b) the case of $T=205$~MeV ($179$~MeV) is denoted by the solid 
(dashed) curve 
. 
}
\label{potential_mu0}
\end{figure}

Figure \ref{potential_mu0} presents the potential surface 
at $\mu =0$. 
Panels (a)-(c) correspond to 
cases of $T/T_{\rm E}=0.73$, $T/T_{\rm E}=0.87,~1$ and 
$T/T_{\rm E}=1.95$, respectively. 
Only one solution $\Omega_0$ 
with $\psi =0$ appears above $T_{\rm E}$, 
while one solution $\Omega_{\rm gr}$ with $\psi=0$ does below $T_{\rm E}$. 
This is consistent with the fact that in Fig.~\ref{theta_Omega_unstable} 
there is only one solution with $\phi =0$ at $\theta=0$. 
At $\mu =0$, thus, the phase $\psi$ is always zero for any $T$. 
This property guarantees the $C$ conservation, but 
does not induce any singularity in the $T$ dependence of physical quantities. 
However, the transition of the ground-state structure 
from the $T<T_{\rm E}$ regime to the $T>T_{\rm E}$ regime 
makes 
$|\Psi |$ singular at the endpoint $(\theta,T)=(\pi,T_{\rm E})$ of 
the RW phase transition and then induces 
a rapid change of $|\Psi |$ even at $\mu =0$.

\section{Summary}
\label{sec:summary}

We have analyzed 
the Roberge-Weiss (RW) mechanism and the RW phase transition, 
using the PNJL model. 
Above $T_{\rm E}$, three ${\mathbb Z}_{3}$ vacua appear alternatively as 
$\theta$ changes from 0 to $2\pi$. 
As a consequence of this RW mechanism, the extended ${\mathbb Z}_{3}$ symmetry 
(the RW periodicity) is preserved, but $C$ symmetry is broken 
at $\theta=\pi/3$ mod $2\pi/3$. 
This is the origin of the RW phase transition. 
As an order parameter of the phase transition, we can select anyone of 
$\theta$-odd quantities; 
a typical one is the phase $\psi$ of the modified Polyakov loop. 
The RW phase transition is of second order 
at the endpoint $(\theta,T)=(\pi/3~{\rm mod}~2\pi/3,T_{\rm E})$. 

The QCD system has ${\mathbb Z}_{3}$ vacua above $T_{\rm E}$, but 
does not below $T_{\rm E}$.  As a consequence of the transition, 
the Polykov-loop $|\Phi|$ has a singular behavior at the endpoint 
$(\theta,T)=(\pi/3~{\rm mod}~2\pi/3,T_{\rm E})$ of 
the RW phase transition. 
The singular behavior induces a rapid change of $|\Phi|$ 
at $\theta =0$, as presented in Fig.~\ref{Psi_all}. 
Thus, the crossover deconfinement transition 
at $\mu =0$, defined by the rapid change of $|\Phi|$, 
is a remnant of the second-order RW phase transition at 
the endpoint $(\theta,T)=(\pi/3~{\rm mod}~2\pi/3,T_{\rm E})$.

Just above $T_{\rm E}$,  
one or two of ${\mathbb Z}_{3}$ vacua emerge at each $\theta$ 
in the PNJL model, while 
all of them appear at each $\theta$ in the RW prediction 
based on perturbation. 
Thus, the RW mechanism is seen 
also in the strong-coupling regime which the PNJL model treats, 
but it is somewhat different from that predicted by Roberge and Weiss 
in the weak-coupling regime.

\bigskip

\noindent
\begin{acknowledgments}
The authors thank M. Matsuzaki and M. Sato for useful discussions. 
H.K. also thanks M. Imachi, H. Yoneyama, K. Takenaga and M. Tachibana for useful discussions. 
This calculation was partially carried out on SX-8 at Research Center for Nuclear physics, Osaka University. 
\end{acknowledgments}


\end{document}